\documentclass[twoside,8pt]{article}
\usepackage{epsfig}

\newcommand{\be}{\begin{equation}}
\newcommand{\ee}{\end{equation}}
\newcommand{\bea}{\begin{eqnarray}}
\newcommand{\eea}{\end{eqnarray}}

\begin{document}

\title{Thermal effects on nuclear symmetry energy with a momentum-dependent effective interaction }

\author{Ch.C. Moustakidis\\
$^{}$ Department of Theoretical Physics, Aristotle University of
Thessaloniki, \\ 54124 Thessaloniki, Greece }

\maketitle

\begin{abstract}
The knowledge of the nuclear symmetry energy  of hot neutron-rich
matter is important for understanding the dynamical evolution of
massive stars and the supernova explosion mechanisms. In
particular, the electron capture  rate on nuclei and/or free
protons in presupernova explosions is especially sensitive to the
symmetry energy at finite temperature. In view of the above, in
the present work we calculate the symmetry energy as a function of
the temperature for various values of the baryon density, by
applying a momentum-dependent effective interaction. In addition
to a previous work, the thermal effects are studied separately
both in the kinetic part and the interaction part of the symmetry
energy. We focus also on the calculations of the mean field
potential, employed extensively in heavy ion reaction research,
both for nuclear and pure neutron matter. The proton fraction and
the electron chemical potential, which are crucial quantities for
representing  the thermal evolution of supernova and neutron
stars, are calculated for various values of the temperature.
Finally, we construct a temperature dependent equation of state of
$\beta$-stable nuclear matter, the basic ingredient for the
evaluation of the neutron star properties.
\end{abstract}

\section{Introduction}
The determination of the nuclear symmetry energy (SE) based on
microscopic and/or phenomenological approaches is of great
interest in nuclear physics as well as in nuclear astrophysics.
For instance, it is important for the study of the structure and
reactions of neutron-rich nuclei, the Type II supernova
explosions,  neutron-star mergers and  the stability of neutron
stars. In addition, the SE is the basic ingredient for the
determination of the proton fraction and electron chemical
potential. The above quantities determine the cooling rate and
neutrino emission flux of protoneutron stars and the possibility
of kaon condensation in dense matter \cite{Bethe-90,Prakash-97}.

Heavy-ion reactions are a unique means to produce in terrestrial
laboratories hot neutron-rich matter similar to those existing in
many astrophysical situations \cite{Bao-Li-06}. Although  the
behavior of the SE for densities below the saturation point still
remains unknown, significant progress has been made only most
recently in constraining the SE at subnormal densities and around
the normal density from the isospin diffusion data in heavy-ion
collisions \cite{Wen05, Bao05}. This has led to a significantly
more refined constraint on neutron-skin thickness of heavy nuclei
\cite{Steiner05, Chen05} and the mass-radius correlation of
neutron stars \cite{Li-06}. For densities above the saturation
point the trend of the SE is model dependent and exhibits
completely different behavior.

Up to now the main part of the calculations concerning the density
dependence of the SE is related with the  cold nuclear matter
($T=0$). However, recently, there is an increasing interest for
the study of the SE and  the properties of  neutron stars at
finite temperature
\cite{Bao-Li-06,Donati-94,Mishra-93,Ccernai-92,Zuo-03,Chen-01,Xu-07-1,Xu-07-2}.
The motivation of the present work is to clarify the effects of
finite temperature on SE  and to find also the appropriate
relations describing that effect. Especially we focus on the
interaction part of the SE, where  so far it has received little
theoretical attention concerning its dependence on the
temperature.

In order to investigate the thermal properties of the SE, we apply
a momentum dependent effective interaction model. In that way, we
are able to study  simultaneously thermal effects not only on the
kinetic part of the symmetry energy but also on the interaction
part. The present model has been introduced by Gale et al.
\cite{Gale-87,Gale-90,Bertsch-88,Prakash-88-1} in order to examine
the influence of momentum-dependent interactions on the momentum
flow of heavy ion collisions. Over the years the model has been
extensively applied in the study not only of the heavy ion
collisions but also in the properties of nuclear matter by a
proper modification \cite{Das-03,Das-07,Bao-Li-04,Chen-05}. A
review analysis of the present model is presented in Refs.
\cite{Prakash-97,Bertsch-88}.

In the present work we study the thermal properties of the nuclear
symmetry energy by applying the above phenomenological model
focusing mainly on the temperature dependence of the kinetic and
interaction part of the SE as well as the total SE. Though it is
well known how the temperature affects the kinetic part of the
symmetry energy \cite{Bao-Li-06,Lee-01,Mekjian-05} the temperature
dependence of the interaction part of the SE has so far received
little theoretical attention. In addition, we determine the
temperature dependence of the proton fraction as well as of the
electron chemical potential. Both of the above quantities are
related with the thermal evaluation of the supernova and the
proton-neutron stars. The single particle potential for the pure
neutron matter and the symmetric nuclear matter, extensively
applied in heavy ion collision research, is also estimated for
various values of the temperature. Finally, we construct the
equation of state (EOS) of $\beta$-stable matter which is the
basic ingredient for  calculations of the neutron star properties.

%The present work aims to be added on the framework of the previous
%studies concerning the effects of temperature on nuclear symmetry
%properties \cite{}.

The plan of the paper is as follows. In Sec.~II the model and the
relative formulae are discussed and analyzed. Results are reported
and  discussed in Sec.~III, while the summary of the work is given
in Sec.~IV.

\section{The model}
The schematic potential model, used in the present work, is
designed  to reproduce the results of the more microscopic
calculations of both nuclear and neutron-rich matter at zero
temperature and can be extended to finite temperature
\cite{Prakash-97}. The energy density of the asymmetric nuclear
matter (ANM) is given by the relation
\begin{equation}
\epsilon(n_n,n_p,T)=\epsilon_{kin}^{n}(n_n,T)+\epsilon_{kin}^{p}(n_p,T)+
V_{int}(n_n,n_p,T), \label{E-D-1}
\end{equation}
where $n_n$ ($n_p$) is the neutron (proton) density and the total
baryon density is $n=n_n+n_p$. The contribution of the kinetic
parts are
\begin{equation}
\epsilon_{kin}^n(n_n,T)+\epsilon_{kin}^p(n_p,T)=2 \int \frac{d^3
k}{(2 \pi)^3}\frac{\hbar^2 k^2}{2m}
\left(f_n(n_n,k,T)+f_p(n_p,k,T) \right), \label{E-K-D-1}
\end{equation}
where $f_{\tau}$, (for $\tau=n,p$) is the Fermi-Dirac distribution
function with the form
\begin{equation}
f_{\tau}(n_{\tau},k,T)=\left[1+\exp\left(\frac{e_{\tau}(n,k,T)-\mu_{\tau}(n,T)}{T}\right)
\right]^{-1}. \label{FD-1}
\end{equation}
The nucleon density $n_{\tau}$ is evaluated from the following
integral
\begin{equation}
n_{\tau}=2 \int \frac{d^3k}{(2\pi)^3}f_{\tau}(n_{\tau},k,T)=2 \int
\frac{d^3k}{(2\pi)^3}\left[1+\exp\left(\frac{e_{\tau}(n,k,T)-\mu_{\tau}(n,T)}{T}\right)\right]^{-1}.
\label{D-1}
\end{equation}
In Eq. (\ref{FD-1}), $e_{\tau}(n,k,T)$ is the single particle
energy (SPE) and $\mu_{\tau}(n,T)$ stands for the chemical
potential of each species. The SPE has the form
\begin{equation}
e_{\tau}(n,k,T)=\frac{\hbar^2k^2}{2m}+U_{\tau}(n,k,T),
\label{esp-1}
\end{equation}
where the single particle potential $U_{\tau}(n,k,T)$, is obtained
by  differentiating $V_{int}$ i.e. $U_{\tau}=\partial
V_{int}(n_n,n_p,T)/\partial n_{\tau}$.
Including the effect of finite-range forces between nucleons, in
order to avoid acausal behavior at high densities, the potential
contribution is parameterized as follows \cite{Prakash-97}
\begin{eqnarray}
V_{int}(n_n,n_p,T)&=&\frac{1}{3}An_0\left[\frac{3}{2}-(\frac{1}{2}+x_0)(1-2x)^2\right]u^2
+\frac{\frac{2}{3}Bn_0\left[\frac{3}{2}-(\frac{1}{2}+x_3)(1-2x)^2\right]u^{\sigma+1}}
{1+\frac{2}{3}B'\left[\frac{3}{2}-(\frac{1}{2}+x_3)(1-2x)^2\right]u^{\sigma-1}}
\nonumber \\ &+& \frac{2}{5}u \sum_{i=1,2}\left[(2C_i+4Z_i) \ 2
\int \frac{d^3k}{(2\pi)^3} g(k,\Lambda_i)(f_n+f_p) \right. \nonumber \\
&+& \left. (C_i-8Z_i) \ 2 \int \frac{d^3k}{(2\pi)^3}
g(k,\Lambda_i)(f_n(1-x)+f_px) \right], \label{V-all}
\end{eqnarray}
where $x=n_p/n$ is the proton fraction and $u=n/n_0$, with $n_0$
denoting the equilibrium symmetric nuclear matter density
$n_0=0.16$ fm$^{-3}$. The constants $A$, $B$, $\sigma$, $C_1$,
$C_2$ and $B'$, which enter in the description of symmetric
nuclear matter and the additional parameters $x_0$, $x_3$, $Z_1$,
and $Z_2$, used to determine the properties of asymmetric nuclear
matter, are treated as parameters  constrained by empirical
knowledge \cite{Prakash-97}. The function $g(k,\Lambda_i)$
suitably chosen to simulate finite range effects is  of the
following form
\begin{equation}
g(k,\Lambda_i)=\left[1+\left(\frac{k}{\Lambda_{i}}\right)^2
\right]^{-1}, \label{G-1}
\end{equation}
where the finite range parameters are $\Lambda_1=1.5 k_F^{0}$ and
$\Lambda_2=3 k_F^{0}$ and $k_F^0$ is the Fermi momentum at the
saturation point $n_0$.

The entropy density $s_{\tau}(n,T)$ required for the calculations
of the total pressure and for the EOS, has the same functional
form as that of a non interacting gas system, that is
\begin{equation}
s_{\tau}(n,T)=-2\int \frac{d^3k}{(2\pi)^3}\left[f_{\tau} \ln
f_{\tau}+(1-f_{\tau}) \ln(1-f_{\tau})\right].  \label{s-den-1}
\end{equation}
The ratio entropy/baryon is given by
$S_{\tau}(n,T)=s_{\tau}(n,T)/n$. The baryon pressure $P_b(n,T)$,
needed to construct the EOS,  is given by
\begin{equation}
P_b(n,T)=T\sum_{\tau=p,n}s_{\tau}(n,T)+\sum_{\tau=p,n}n_{\tau}\mu_{\tau}(n,T)-\epsilon_{anm}(n,T).
\label{P-1}
\end{equation}
Finally, the total energy density and pressure of charge neutral
and chemically equilibrium nuclear matter are
\begin{equation}
\epsilon_{tot}(n,T)=\epsilon_{b}(n,T)+\sum_{l=e^-,\mu^-}
\epsilon_l(n,T), \label{e-total}
\end{equation}
\begin{equation}
P_{tot}(n,T)=P_{b}(n,T)+\sum_{l=e^-,\mu^-} P_l(n,T).
\label{P-total}
\end{equation}
The leptons (electrons and muons) originating from the condition
of the $\beta$-stable matter are  considered as non-interacting
Fermi gases.

The above analysis holds in general for the asymmetric nuclear
matter. Below, in order to calculate the thermal effect on the SE,
we will focus our study on two cases, i.e. the symmetric nuclear
matter (SNM) and the pure neutron matter (PNM).

%%%%%%%%%%%%%%%%%%%%%%%%%%%%%%%%%%%%%%%%%%%%%%%%%%
\subsection{Symmetric nuclear matter}
%%%%%%%%%%%%%%%%%%%%%%%%%%%%%%%%%%%%%%%%%%%%%%%%%%
The energy density of  SNM is given by Eqs. (\ref{E-D-1}) and
(\ref{V-all}) by setting $x=1/2$, that is \cite{Prakash-97}
\begin{eqnarray}
\epsilon_{snm}(n,T)&=&2\int
\frac{d^3k}{(2\pi)^3}\frac{\hbar^2k^2}{2m}f_n +2\int
\frac{d^3k}{(2\pi)^3}\frac{\hbar^2k^2}{2m}f_p
+\frac{1}{2}An_0u^2+\frac{Bn_0u^{\sigma+1}}{1+B'u^{\sigma-1}}\nonumber\\
&+&u\sum_{i=1,2}C_i \ 2 \int \frac{d^3k}{(2\pi)^3}
g(k,\Lambda_i)f_n+ u\sum_{i=1,2}C_i \ 2 \int \frac{d^3k}{(2\pi)^3}
g(k,\Lambda_i)f_p . \label{ensnm-1}
\end{eqnarray}
In addition, the single particle potential $U_{snm}^{\tau}(n,k,T)$
in the case of SNM, defined from the relation
$U_{snm}^{\tau}=\partial V_{snm}/\partial n_{\tau}$, is easily
calculated and given by
\begin{equation}
U_{snm}^{\tau}(n,k,T)=\tilde{U}_{snm}^{\tau}(n,T)+
u\sum_{i=1,2}C_i\left[1+\left(\frac{k}{\Lambda_i}\right)^2\right]^{-1}.
\label{Us-1}
\end{equation}
It is obvious from Eq.~(\ref{Us-1}) that $U_{snm}^{\tau}(n,k,T)$
is separated in two terms. The first one corresponds to the
momentum independent part, while the second one corresponds to the
momentum dependent one. The term $\tilde{U}_{snm}^{\tau}(n,T)$ has
the following form
\begin{eqnarray}
\tilde{U}_{snm}^{\tau}(n,T)&=&
Au+\frac{Bu^{\sigma}(\sigma+1+2B'u^{\sigma-1})}{(1+B'u^{\sigma-1})^2}
\nonumber\\
&+&\frac{2}{n_0} \sum_{i=1,2}C_i 2 \int
\frac{d^3k}{(2\pi)^3}\left[1+\left(\frac{k}{\Lambda_i}\right)^2\right]^{-1}
f_{\tau}, \qquad \tau=p,n.  \label{Us-2}
\end{eqnarray}

At zero temperature ($T=0$), where $\theta(k_{F_{\tau}}-k)$, the
integrals in Eqs.~(\ref{ensnm-1}) and (\ref{Us-2}) are calculated
analytically (see Appendix A for more details).

%%%%%%%%%%%%%%%%%%%%%%%%%%%%%%%%%%%%%%%%%%%%%%%%%%
\subsection{Pure nuclear matter}
%%%%%%%%%%%%%%%%%%%%%%%%%%%%%%%%%%%%%%%%%%%%%%%%%%
The energy density of  PNM is given by Eqs. (\ref{E-D-1}) and
(\ref{V-all}) by setting $x=0$ and $f_p=0$, that is
\cite{Prakash-97}
\begin{eqnarray}
\epsilon_{pnm}(n,T)&=&2\int
\frac{d^3k}{(2\pi)^3}\frac{\hbar^2k^2}{2m}f_n+\frac{1}{3}An_0(1-x_0)u^2+
\frac{\frac{2}{3}Bn_0(1-x_3)u^{\sigma+1}}{1+\frac{2}{3}B'(1-x_3)u^{\sigma-1}}\nonumber
\\
&+&\frac{2}{5}u\sum_{i=1,2}(3C_i-4Z_i) \ 2 \int
\frac{d^3k}{(2\pi)^3} g(k,\Lambda_i)f_n .\label{epnm-1}
\end{eqnarray}
The single particle potential $U_{pnm}^{n}(n,k,T)$ in the case of
PNM is defined from the relation $U_{pnm}^n=\partial
V_{pnm}/\partial n_n$ is written as
\begin{equation}
U_{pnm}^n(n,k,T)=\tilde{U}_{pnm}^n(n,T)+
\frac{2}{5}u\sum_{i=1,2}(3C_i-4Z_i)\left[1+\left(\frac{k}{\Lambda_i}\right)^2\right]^{-1}.
\label{Un-1}
\end{equation}
The momentum-independent part is
\begin{eqnarray}
\tilde{U}_{pnm}^n(n,T)&=&\frac{2}{3}A(1-x_0)u+\frac{\frac{2}{3}B(1-x_3)u^{\sigma}}{[1+\frac{2}{3}B'(1-x_3)u^{\sigma-1}]^2}
\left((\sigma+1)+\frac{4}{3}B'(1-x_3)u^{\sigma-1}\right)\nonumber\\
&+&\frac{2}{5n_0} \sum_{i=1,2}(3C_i-4Z_i) 2 \int
\frac{d^3k}{(2\pi)^3}\left[1+\left(\frac{k}{\Lambda_i}\right)^2\right]^{-1}
f_n. \label{Un-2}
\end{eqnarray}

The integrals in Eqs.~(\ref{epnm-1}) and (\ref{Un-2}), similarly
to the case of SNM, at $T=0$ are calculated analytically (see
Appendix A for more details).

%%%%%%%%%%%%%%%%%%%%%%%%%%%%%%%%
\subsection{Asymmetric nuclear matter-Nuclear symmetry energy}
The energy density of ANM at density $n$ and temperature $T$, in a
good approximation, is expressed as
\begin{equation}
\epsilon_{anm}(n,T,x)=\epsilon_{snm}(n,T,x=1/2)+\epsilon_{sym}(n,T,x),
\label{e-asm-1}
\end{equation}
where
\begin{equation}
\epsilon_{sym}(n,T,x)=n(1-2x)^2 E_{sym}^{tot}(n,T)=n (1-2x)^2
\left(E_{sym}^{kin}(n,T)+E_{sym}^{int}(n,T)\right).
\label{e-sym-1}
\end{equation}
In Eq.~(\ref{e-sym-1}) the nuclear symmetry energy
$E_{sym}^{tot}(n,T)$ is separated in   two parts corresponding to
the kinetic contribution $E_{sym}^{kin}(n,T)$ and the interaction
contribution $E_{sym}^{int}(n,T)$. In the present work we will
concentrate on the systematic study of the thermal properties of
the above two quantities.

From Eqs.~(\ref{e-asm-1}) and (\ref{e-sym-1}) and setting $x=0$ we
obtain that the nuclear symmetry energy $E_{sym}^{tot}(n,T)$ is
given by
\begin{equation}
E_{sym}^{tot}(n,T)=\frac{1}{n}\left(\epsilon_{pnm}(n,T)-\epsilon_{snm}(n,T)
\right). \label{Esym-d-1}
\end{equation}
Thus, from Eqs.~(\ref{ensnm-1}) and (\ref{epnm-1}) and by a
suitable choice of  the parameters $x_0$, $x_3$, $Z_1$ and $Z_2$,
we can obtain different forms for the density dependence of the
symmetry energy $E_{sym}^{tot}(n,T)$. It is well known that the
need to explore different forms for $E_{sym}^{tot}(n,T)$ stems
from the uncertain behavior at high density \cite{Prakash-97}. In
the present work, since we are interested mainly in the study of
thermal effects on the SE, we choose a specific form of the SE
enabling us  to reproduce accurately  the results of many other
theoretical studies \cite{Lee-97}. According to this choice the
SE, at $T=0$, is expressed as
\begin{equation}
E_{sym}^{tot}(n,T=0)= \underbrace{13
u^{2/3}}_{Kinetic}+\underbrace{17
F(u)}_{Interaction}=\underbrace{13
u^{2/3}}_{Kinetic}+\underbrace{17 u}_{Interaction},\label{Esym-3}
\end{equation}
where the contributions of the kinetic  and the interaction term
are separated clearly. The parameters $x_0$, $x_3$, $Z_1$ and
$Z_2$ are chosen in order that Eq.~(\ref{Esym-d-1}), for $T=0$, to
reproduce the results of Eq.~(\ref{Esym-3}). In addition, the
parameters $A$, $B$, $\sigma$, $C_1$, $C_2$ and $B'$ are
determined  in order that $E(n=n_0)-mc^2=-16$ {\rm MeV},
$n_0=0.16$ fm$^{-3}$, and the incompressibility to be $K_0=240$
{\rm MeV}.

The single particle potential $U_{anm}^{\tau}(n,k,T)$, in the case
of ANM  defined from the relation $U_{anm}^{\tau}=\partial
V_{anm}/\partial n_{\tau}$, is written as
\begin{equation}
U_{anm}^{\tau}(n,k,T)=U_{snm}^{\tau}(n,k,T)+\frac{\partial
V_{sym}}{\partial n_{\tau}}=U_{snm}^{\tau}(n,k,T)+
U_{sym}^{\tau}(n,T,x),\label{Uasnm-1}
\end{equation}
where
\begin{equation}
V_{sym}(n,T,x)=(1-2x)^2n E_{sym}^{int}(n,T). \label{V-sym-1}
\end{equation}
It is easy to find that the term $U_{sym}^{\tau}(n,T)$, in the
case of $T=0$ and by applying  expression (\ref{Esym-3}), is given
by (see also ref. \cite{Bao-97})
\begin{equation}
U_{sym}^{\tau}(n,T,x)=\pm 34 u (1-2x), \qquad  \label{V-anm-1}
\end{equation}
where $+$ and$-$ stand for neutrons and protons respectively.
In the general case where thermal effects are included in our
calculations, the  $E_{sym}^{int}(n,T)$ takes the form
\begin{equation}
E_{sym}^{int}(n,T)=a u^b, \label{E-int-T}
\end{equation}
where $a$ and $b$ are temperature dependent constants (see
Eq.~(\ref{Esym-pot-fit}) on Sec.~III). Thus, after some algebra,
we get in a good approximation, the relation
\begin{equation}
U_{sym}^{\tau}(n,T,x)\simeq \pm 2 a u^b (1-2x). \label{V-anm-T}
\end{equation}
The above relation is needed for the calculation of the single
particle energy $e_{\tau}(n,k,T)$ in the $\beta$-stable matter and
afterwards for the calculation of the Fermi-Dirac function
$f_{\tau}(n,T)$ which is the basic ingredient for the
determination of the entropy density $s_{\tau}(n,T)$.

\subsection{Proton fraction-Electron chemical potential}
The key quantity for the determination of the equation of state in
$\beta$-stable matter is the proton fraction $x$, which is a basic
ingredient of  Eq.~(\ref{e-sym-1}). In $\beta$-stable matter the
processes \cite{Prakash-94}
\begin{equation}
n \longrightarrow p+e^{-}+\bar{\nu}_e, \qquad \qquad p +e^{-}
\longrightarrow n+ \nu_e,
\end{equation}
take place simultaneously. We assume that neutrinos generated in
these reactions have left the system. This implies that
\begin{equation}
\hat{\mu}=\mu_n-\mu_p=\mu_e ,\label{chem-1}
\end{equation}
where $\mu_n,\mu_p$ and $\mu_e$ are the chemical potentials of the
neutron, proton and electron respectively. Given the total energy
density $\epsilon \equiv \epsilon(n_n,n_p)$, the neutron and
proton chemical potentials can be defined as
\begin{equation}
\mu_n=\frac{\partial \epsilon}{\partial n_n}|_{n_p}, \qquad \qquad
\mu_p=\frac{\partial \epsilon }{\partial n_p}|_{n_n} .
\label{chem-2}
\end{equation}

Hence we can show that
\begin{equation}
\hat{\mu}=\mu_n-\mu_p=-\frac{\partial \epsilon /n}{\partial x}|_n=
 -\frac{\partial E}{\partial x}|_n . \label{chem-3}
\end{equation}

In $\beta$ equilibrium one has
\begin{equation}
\frac{\partial E}{\partial x}=\frac{\partial}{\partial
x}\left(E_b(n,x)+E_e(x)\right)=0 , \label{b-equil-1}
\end{equation}
where $E_b(n,x)$ the energy per baryon and $E_e(x)$ the electron
energy. The charge condition implies that $n_e=n_p=nx$ or
$k_{F_e}=k_{F_p}$. Combining  relations  (\ref{e-asm-1}),
(\ref{e-sym-1})  and (\ref{chem-3}) we get
\begin{equation}
\mu_e(n,T)=\hat{\mu}(n,T)=4(1-2x)E_{sym}^{tot}(n,T) .
\label{chem-4}
\end{equation}
From Eq.~(\ref{chem-4}) it is obvious that the proton fraction $x$
is not only a function of the baryon density $n$ but, in addition,
depends on the temperature $T$ i.e.  $x=x(n,T)$.

For relativistic non-degenerate free electrons we have
\begin{equation}
n_e=xn=\frac{2}{(2\pi^3)}\int\frac{d^3k}
{1+\exp\left[\frac{\sqrt{\hbar^2k^2c^2+m_e^2c^4}-\mu_e(n,T)}{T}\right]}.
\label{ele-frac-1}
\end{equation}
Or, using Eq.~(\ref{chem-4}) and performing the angular
integration we get
\begin{equation}
n_e=xn=\frac{1}{\pi^2}\int_0^{\infty} \frac{k^2 dk}
{1+\exp\left[\frac{\sqrt{\hbar^2k^2c^2+m_e^2c^4}-4(1-2x)E_{sym}^{tot}(n,T)}{T}\right]}.
\label{ele-frac-2}
\end{equation}
Eq.~(\ref{ele-frac-2}) determines the equilibrium electron
(proton) fraction $x(n,T)$ since the density and momentum
dependent symmetry energy $E_{sym}^{tot}(n,T)$ is known.

%%%%%%%%%%%%%%%%%%%%%%%%%%%%%%%%%%%%
\subsection{Calculations recipe}
%%%%%%%%%%%%%%%%%%%%%%%%%%%%%%%%%%%%
We focus our attention on the calculation of the
$E_{sym}^{tot}(n,T)$ with the help of Eq. (\ref{Esym-d-1}). Thus,
one has to calculate first the energy densities in pure and in
symmetric nuclear matter as a function of the density $n$ and for
fixed values of  temperature $T$. As an example of the
calculations procedure at finite temperature (the results for
$T=0$ are included in the Appendix A), we consider the case of
pure neutron matter. The procedure is similar in the case of
symmetric nuclear matter (see Ref. \cite{Prakash-97}).

The outline of our approach is the following: For a fixed neutron
density $n_n$ and temperature $T$, Eq.~(\ref{D-1}) may be solved
iteratively in order to calculate the variable
\begin{equation}
\eta(n;T)=\frac{\mu_{\tau}(n;T)-\tilde{U}(n;T)}{T}. \label{eta-1}
\end{equation}

The knowledge of $\eta(n,T)$ allows the last term in Eq.
(\ref{Un-2})  to be evaluated, yielding $\tilde{U}(n;T)$ which may
then be used to infer the chemical potential from
\begin{equation}
\mu_{\tau}(n;T)=T\eta(n;T)+\tilde{U}(n;T), \label{Chem-1}
\end{equation}
required as an input to the calculation of the single particle
spectrum $e_{\tau}(n,k,T)$ in Eq.~(\ref{esp-1}). Using
$e_{\tau}(n,k;T)$, the energy density in Eq.~(\ref{epnm-1})  is
evaluated.

%%%%%%%%%%%%%%%%%%%%%%%%%%%%%%%%%%%%%%%%%%%%%
\section{Results and Discussion}
%%%%%%%%%%%%%%%%%%%%%%%%%%%%%%%%%%%%%%%%%%%%%
According to our calculation recipe, given in the previous
subsection, we calculate the energy densities of  PNM and  SNM as
functions of the density, for various values of the temperature
$T$. As a second step, we calculate the $E_{sym}^{tot}(n,T)$ from
Eq. (\ref{Esym-d-1}). The knowledge of $E_{sym}^{tot}(n,T)$ is
required for the evaluation of the proton fraction $x$ from Eq.
(\ref{ele-frac-2}) as well as for the electron chemical potential
$\mu_e=\hat{\mu}$ from Eq. (\ref{chem-4}). Finally from Eqs.
(\ref{P-1}), (\ref{e-total}) and (\ref{P-total}) we construct the
EOS of $\beta$-stable matter for various values of the temperature
$T$. It is worth  pointing out that in the present work we do not
include the muon case,  since we restrict  ourselves  mainly on
the temperature dependent behavior of the SE. According to our
plan, in future work we will extend the treatment to include also
the muon case in order to study the detailed composition and the
thermal properties of neutron-rich matter with applications in
neutron star structure and thermal evaluation.

In Fig.~1 we check the validity of approximation (\ref{e-asm-1}).
We plot the difference $E(n,T,x)-E(n,T,x=1/2)$ as a function of
$(1-2x)^2$ at temperature $T=0$, $T=20$ and $T=50$ MeV for three
baryon number fractions i.e. $u=1$, $u=2$ and $u=3$. It is seen
that an almost linear relation holds between
$E(n,T,x)-E(n,T,x=1/2)$ and $(1-2x)^2$, even closer to the case of
pure neutron matter ($x=0$), indicating the validity of
approximation (\ref{e-asm-1}).

In Fig.~2 we indicate the behavior of the SE as a function of the
temperature $T$ for various fixed values of the baryon density
$n$. More precisely, in any case, we plot $E_{sym}^{tot}(T;n)$, as
well as $E_{sym}^{kin}(T;n)$ and $E_{sym}^{int}(T;n)$ as a
function of $T$ for $n=0.1, 0.2, 0.3, 0.5$ fm$^{-3}$. The most
striking feature of the above analysis is a decrease of the SE
(total, kinetic and interaction part) by increasing the
temperature. This is consistent with the predictions of
microscopic and/or phenomenological theories
\cite{Bao-Li-06,Chen-01,Xu-07-1}

In order to illustrate further the dependence of the symmetry
energy on the temperature and to find the quantitative
characteristic on this dependence, the values of $E_{sym}(T;n)$
for various values of the density $n$ are  derived with the
least-squares fit method and found to take the general form
\begin{equation}
E_{sym}(T;n)=\frac{A}{1+(T/T_0)^{c}}+B. \label{Logistic-fit}
\end{equation}
The values of the density dependent parameters $A$, $B$, $T_0$ and
$c$, for $E_{sym}^{tot}(T;n)$, $E_{sym}^{kin}(T;n)$ and
$E_{sym}^{int}(T;n)$ for $n=0.1, 0.3, 0.5$ fm$^{-3}$ are presented
in Table~1. It is easy to find that in the case of low temperature
limit ($T/T_0\ll 1$) all kinds of the symmetry energy decrease
approximately according to $E_{sym}(T;n) \propto C_1-C_2T^2$
(where $C_1$  and $C_2$ density dependent constants). In the high
density limit ($T/T_0\gg 1$) the symmetry energy decreases
approximately according to $E_{sym}(T;n) \propto C_3 T^{-2}+C_4$
(where also $C_3$  and $C_4$ are density dependents constants). It
is noted that the same behavior holds for  $E_{sym}^{tot}(T;n)$ as
well as for $E_{sym}^{kin}(T;n)$ and $E_{sym}^{int}(T;n)$. This
behavior is well expected for  the kinetic part of the symmetry
energy (see also Ref. \cite{Bao-Li-06,Mekjian-05}), where
analytical calculations are possible (see the prove in Appendix
B). From the above study, it is concluded that there is a similar
temperature dependence both for the kinetic and the interaction
part of the symmetry energy and consequently for the total
symmetry energy, in the case of momentum dependent interaction.
Recently, the temperature dependence of the kinetic and
interaction part of the SE has been studied and illustrated in
Ref. \cite{Xu-07-1}. The results of the present work agree with
those of Ref. \cite{Xu-07-1}  although different models have been
employed  to evaluate  SE.

%%%%%%%%%%%%%%%%%%%%%%%%%%%%%%%%%%%%%%%%%%%%%%%%%%%%%%%%%%%%%%%%%

In Fig.~3, we plot  $E_{sym}^{tot}(T;n)$ as a function of
temperature for various low values of the baryon density. In the
same figure we also include  experimental data of the measured
temperature dependent symmetry energy from Texas A$\&$M University
(TAMU)\cite{Shetty-06} and the INDRA-ALADIN Collaboration at GSI
\cite{Fevre-05}. The comparison then allows to estimate the
required density of the fragment-emitting of the experiments. As
pointed out by Li et.al. \cite{Bao-Li-06} the experimentally
observed evolution of the SE is mainly due to the change in
density rather than temperature.

%%%%%%%%%%%%%%%%%%%%%%%%%%%%%%%%%%%%%%%%%%%%%%%%%%%%%%%%%%%%%%%%%%%
Fig.~4 illustrates the behavior of the $E_{sym}^{total}(n;T)$ (a),
$E_{sym}^{kin}(n;T)$ (b), $E_{sym}^{int}(n;T)$ (c), as a function
of the baryon density $n$ for various fixed values of the
temperature $T$. The case $T=0$ corresponds to the fundamental
expression of the present work i.e.
\begin{equation}
E_{sym}^{tot}(u;T=0)=13 u^{2/3}+17u. \label{E-u-1}
\end{equation}
In any case, the trends of the various parts of the symmetry
energy are similar. An increase in the temperature leads just to a
shift to  lower values for the symmetry energy. It is worth
pointing out that, the maximum decrease of  $E_{sym}^{tot}(n;T)$,
in the area  under study (for $T=0$ MeV up to $T=50$ MeV), is
between $40\%$ (for $n=0.1$ fm$^{-3}$) and $4\%$ (for
$n=1$fm$^{-3}$). Correspondingly, the decrease  of
$E_{sym}^{kin}(n;T)$ is between $57\%$ (for $n=0.1$ fm$^{-3}$) and
$5\%$ (for $n=1$ fm$^{-3}$) and of the $E_{sym}^{int}(n;T)$ is
between $22\%$ (for $n=0.1$ fm$^{-3}$) and $5\%$ (for $n=1$
fm$^{-3}$). It is obvious that the thermal effects are more
pronounced on the kinetic part  than in the interaction part of
the symmetry energy and in addition, more pronounced in  lower
values of the baryon density.

The total symmetry energy $E_{sym}^{tot}(u;T)$, for various values
of the temperature $T$, was derived with the least-squares fit on
the numerical results taken from Eq.~(\ref{Esym-d-1}) and has the
form
\begin{eqnarray}
E_{sym}^{tot}(u;T=5)&=&1.676+29.711u-2.110u^2+0.275u^3-0.015u^4,
\nonumber \\
E_{sym}^{tot}(u;T=10)&=&-0.118+30.863u-2.455u^2+0.325u^3-0.017u^4, \nonumber \\
E_{sym}(u;T=20)&=&-1.910+29.470u-1.466u^2+0.120u^3-0.004u^4,
\nonumber \\
E_{sym}^{tot}(u;T=50)&=&0.099+18.172u+2.9u^2-0.548u^3+0.033u^4.
 \label{Esym-T-fit}
\end{eqnarray}

It is also useful to record some relations for
$E_{sym}^{tot}(u;T)$  derived  by least-squares fit on the
numerical results, in the case where SE is parameterized in a way
similar to that one holding for $T=0$. In that case, the
parametrization is the following (the case $E_{sym}^{tot}(u;T=0)$
is included also for comparison)
\begin{eqnarray}
E_{sym}^{tot}(u;T=0)&=&13u^{2/3}+17u, \nonumber\\
 E_{sym}^{tot}(u;T=5)&=&E_{sym}^{tot}(u;T=0)-0.374 \ u^{-0.956},
\nonumber \\
E_{sym}^{tot}(u;T=10)&=& E_{sym}^{tot}(u;T=0)-1.235 \ u^{-0.804}, \nonumber \\
E_{sym}(u;T=20)&=&E_{sym}^{tot}(u;T=0)-3.420 \ u^{-0.520},
\nonumber \\
E_{sym}^{tot}(u;T=50)&=&E_{sym}^{tot}(u;T=0)-9.300 \ u^{-0.097}.
\label{Esym-T-fit-2}
\end{eqnarray}
From Eq.~(\ref{Esym-T-fit-2}), the decrease of the SE as a result
of increasing $T$, is evident.

The interaction part of the symmetry energy $E_{sym}^{int}(u;T)$
for various values of the temperature $T$ was derived by a
least-squares fit on the numerical results taken from
Eqs.~(\ref{e-sym-1}) and  (\ref{Esym-d-1}) and has the form
\begin{eqnarray}
E_{sym}^{int}(u;T=5)&=&17.041 \ u^{0.997}, \nonumber \\
E_{sym}^{int}(u;T=10)&=&16.782 \ u^{1.005}, \nonumber \\
E_{sym}^{int}(u;T=20)&=&16.022 \ u^{1.028}, \nonumber \\
E_{sym}^{int}(u;T=50)&=&13.404 \ u^{1.104}.  \label{Esym-pot-fit}
\end{eqnarray}

Similarly, for the kinetic part of the symmetry energy
$E_{sym}^{kin}(u;T)$ we obtain
\begin{eqnarray}
E_{sym}^{kin}(u;T=5)&=&12.856 \ u^{0.674}, \nonumber \\
E_{sym}^{kin}(u;T=10)&=&12.504 \  u^{0.691}, \nonumber \\
E_{sym}^{kin}(u;T=20)&=&11.518 \ u^{0.736}, \nonumber  \\
E_{sym}^{kin}(u;T=50)&=&8.577 \ u^{0.891}.  \label{Esym-kin-fit}
\end{eqnarray}

%%%%%%%%%%%%%%%%%%%%%%%%%%%%%%%%%%%%%%%%%%%%%%%%%%%%%%%%%%%%%%%%%%%%%%%%%

In Fig.~5 we plot the total energy per particle of the PNM (a) and
of the SNM as a function of the density for various values of the
temperature. In both cases it is concluded that the thermal
effects become more pronounced when $T>10$ MeV and for baryon
densities $n<0.5$ fm$^{-3}$.
%%%%%%%%%%%%%%%%%%%%%%%%%%%%%%%%%%%%%%%%%%%%%%%%%%%%%%%%%%%%%%%%%%%%

Fig.~6 displays the single particle potential $U_{pnm}(n,T,k)$ of
the PNM as a function of the momentum $k$ for various values of
the density $n$ and temperature $T$. An increase of  $T$ leads to
corresponding increase of the values of the $U_{pnm}(n,T,k)$, an
effect, expected to be more pronounced for lower values of the
baryon density ($n=0.1$ fm$^{-3}$) compared to highest ($n=0.5$
fm$^{-3}$). The same trend holds also for the single particle
potential $U_{snm}(n,T,k)$ of the SNM plotted in Fig.~7. Observing
Figs.~6 and ~7 one might expect  that the change of $T$ will
affect slightly the nucleons with high momentum $k$. This could be
seen by plotting the single particle energy $e_{\tau}(n,k,T)$ (see
Eq.~(\ref{esp-1})) as a function of $k$. However, the above effect
cannot be seen in the present work, where we plot just the single
particle potential $U^{\tau}(n,k,T)$ as a function of $k$.
%%%%%%%%%%%%%%%%%%%%%%%%%%%%%%%%%%%%%%%%%%%%%%%%%%%%%

In Fig.~8 we display the single particle potential of neutron
$U^n(n,T,k)$ (Fig.~(a),(b)) and proton $U^p(n,T,k)$
(Fig.~(c),(d)), in $\beta$-stable matter, as a function of the
momentum $k$ for various values of the temperature $T$  for
$n=0.1$ and $n=0.5$ fm$^{-3}$. The potential $U^{\tau}(n,T,k)$ is
evaluated according to Eq.~(\ref{Uasnm-1}. The most striking
feature of Fig.~8 is the reduced thermal effect for high values of
the baryon density, especially in the case of the neutron single
particle potential. In the case of the proton, thermal effects are
more pronounced.

%%%%%%%%%%%%%%%%%%%%%%%%%%%%%%%%%%%%%%%%%%%%%%%%%%%%%%%%%%%%%%%%%%%%%%%%%
In Fig.~9(a) the proton fraction $x$ is displayed, calculated from
Eq. (\ref{ele-frac-2}) as a function of $n$ for various values of
$T$. Thermal effects increase the value of $x$ between $57\%$ (for
$n=0.1$ fm$^{-3}$) and $2\%$ (for $n=1$ fm$^{-3}$). This effect is
directly related with the dependence of $x$ on the symmetry
energy. As  discussed previously, the temperature influences
slightly the symmetry energy at high values of the density and
consequently this is reflected in the values of $x$. It is
stressed that $x$ depends on $T$ in two ways, as one can see from
Eq. (\ref{ele-frac-2}). That is, it depends directly on $T$ due
the Dirac-Fermi distribution and also depends on the symmetry
energy which is also temperature dependent.
%%%%%%%%%%%%%%%%%%%%%%%%%%%%%%%%%%%%%%%%%%%%%%%%%%%%%%%%

In Fig.~9(b) we present the electron chemical potential $\mu_e$ as
a function of the density $n$ for various $T$. An increase of $T$
decreases  $\mu_e$. The effect is more pronounced when $T>20$ {\rm
MeV}. We mention that the rate of electron capture on both free
and bound protons depends in a very sensitive way on the
difference $\hat{\mu}=\mu_n-\mu_p=\mu_e$ between neutron and
proton chemical potentials \cite{Donati-94}. Larger values of
$\hat{\mu}=\mu_e$ inhibit  the neutronization process, since it
becomes more difficult to transform a proton into a neutron.
%%%%%%%%%%%%%%%%%%%%%%%%%%%%%%%%%%%%%%%%%%%%%%%%%%%%%%%%%%%%%%%%%

Finally, in Fig.~10 we present the equation of state of beta
stable matter constructed by applying the present
momentum-dependent interaction model, for various values of the
temperature  $T$. It is obvious that the thermal effects are
enhanced  when $T>20$ {\rm MeV}. The above EOS is very important
for the calculation of the neutron stars properties and also in
combination with the calculated proton fraction and electron
chemical potentials for the thermal evaluation of the neutron
stars.

%%%%%%%%%%%%%%%%%%%%%%%%%%%%%%%%%%%%%%%%%
\section{Summary}
%%%%%%%%%%%%%%%%%%%%%%%%%%%%%%%%%%%%%%%%%
The knowledge of the nuclear symmetry energy  of hot neutron-rich
matter is important for understanding the dynamical evolution of
massive stars and the supernova explosion mechanisms. In view of
the above statement, we investigate, in the present work,  the
thermal effects on the nuclear symmetry energy.  In order to
perform the above investigation we  apply a model with a
momentum-dependent effective interaction. In that way, we are able
to study the thermal  effect not only on the kinetic part of the
symmetry energy but also on the interaction part which, in turn,
due to a momentum dependence, is affected by the variation of the
temperature.  It is concluded that, in general, by increasing $T$
we obtain a decreasing SE. Our finding that both kinetic and
interaction parts exhibit the same trend both for low and high
values of the temperature  is an interesting result. Analytical
relations, derived by the method of least squares fit are given
also for the above quantities. Temperature effects on the pure
neutron matter and also on symmetric nuclear matter are also
investigated and presented. The single particle potential of
proton and neutron is of interest in heavy ions collisions
experiments, is calculated also for pure neutron matter, symmetric
nuclear matter and $\beta$-stable matter for various values of the
baryon density and fixed values of T. It is concluded that thermal
effects are more pronounced for low values of the density $n$,
where for high values of $n$ the effects are almost negligible.
Quantities, which are of great interest for the thermal evaluation
of supernova and neutron stars, i.e. the proton fraction
$x=x(n,T)$ and the electron chemical potential $\mu_e=\mu_e(n,T)$,
are calculated and their temperature and density dependence is
investigated. Thermal effects are larger for low values of the
density and high values of T.

%%%%%%%%%%%%%%%%%%%%%%%%%%%%%%%%%%%%%%%%
\section*{Appendix A}
%%%%%%%%%%%%%%%%%%%%%%%%%%%%%%%%%%%%%%%%
The energy density of the SNM as well as of the PNM, at zero
temperature are easily calculated from Eqs. (\ref{ensnm-1}) and
(\ref{epnm-1}) respectively  by setting
$f_{\tau}=\theta(k_{F_{\tau}}-k)$ (where $\theta(k_{F_{\tau}}-k)$
is the {\it theta} function and $k_{F_{\tau}}$ is the Fermi
momentum of the nucleon $\tau$) and takes the following forms
\begin{eqnarray}
\epsilon_{snm}(n,k;T=0)&=&\frac{3}{5}E_F^0n_0u^{5/3}+
\frac{1}{2}An_0u^2+\frac{Bn_0u^{\sigma+1}}{1+B'u^{\sigma-1}}\nonumber\\
&+& 3n_0u
\sum_{i=1,2}C_i\left(\frac{\Lambda_i}{k_F^0}\right)^3\left(\frac{u^{1/3}}{\frac{\Lambda_i}{k_F^0}}-
\tan^{-1} \frac{u^{1/3}}{\frac{\Lambda_i}{k_F^0}} \right),
\label{ensnm-T0}
\end{eqnarray}
%%%%%%%%%%%%%%%%%%%%%%%%%%%%%%%%%%%%%%%%%%%%%%%%%%%%
\begin{eqnarray}
\epsilon_{pnm}(n,k;T=0)&=&2^{2/3}\frac{3}{5}E_F^0n_0u^{5/3}+
\frac{1}{3}An_0(1-x_0)u^2+
\frac{\frac{2}{3}Bn_0(1-x_3)u^{\sigma+1}}{1+\frac{2}{3}B'(1-x_3)u^{\sigma-1}}
\nonumber\\
&+& \frac{3}{5}n_0u
\sum_{i=1,2}\left(3C_i-4Z_i\right)\left(\frac{\Lambda_i}{k_F^0}\right)^3
\left(\frac{(2u)^{1/3}}{\frac{\Lambda_i}{k_F^0}}- \tan^{-1}
\frac{(2u)^{1/3}}{\frac{\Lambda_i}{k_F^0}} \right),
\label{enpnm-T0}
\end{eqnarray}
where $E_F^0=\hbar^2{k_F^0}^2/2m$ is the Fermi energy of nuclear
matter at the equilibrium density.

\section*{Appendix B}
In order to compare the numerical results obtained from the
kinetic part of the symmetry energy $E_{sym}^{kin}(n,T)$ with
those predicted from analytical calculations, we calculate
$E_{sym}^{total}(n,T)$ in the low and in the hight temperature
limit as follows

\subsubsection*{Low temperature limit}
The kinetic energy per nucleon $E_{kin}^{\tau}(n,T)$ at low
temperature ($T\ll E_F$) has the form
\cite{Goodstein-85,Huang-87,Fetter-03}
\begin{equation}
E_{kin}^{\tau}(n,T)=\frac{3}{5}E_F^{\tau}\left[1+\frac{5}{12}\pi^2\left(\frac{T}{E_F^{\tau}}\right)^2
\right], \label{Ekin-1}
\end{equation}
where $E_F^{\tau}=(\hbar
k_F^{\tau})^2/2m=\hbar^2(3\pi^2n_{\tau})^{2/3}/2m$.
Considering that $\delta=1-2x=(n_n-n_p)/(n_n+n_p)$ after some
algebra we found that the $E_{kin}(n,T,\delta)$ of a two-component
Fermi gas  has the form
\begin{eqnarray}
E_{kin}(n,\delta,T)&=&\frac{\langle E_F \rangle}{2}\left((1+\delta)^{5/3}+(1-\delta)^{5/3}\right)\nonumber\\
&+&\frac{3}{10}\frac{1}{\langle E_F \rangle} \left(\frac{\pi}{2}T
\right)^2 \left((1+\delta)^{1/3}+(1-\delta)^{1/3}\right),
\label{Ek-1}
\end{eqnarray}
where $\langle E_F \rangle=3/5E_F^0$.  Expanding expression
(\ref{Ek-1}) around the symmetric point $\delta=0$ or $x=1/2$ the
kinetic energy takes the approximated form
\begin{equation}
E_{kin}(n,T)=\langle E_F \rangle+\frac{3}{20}\frac{\pi^2}{\langle
E_F \rangle}T^2+ (1-2x)^2 \underbrace{\left(\frac{5}{9}\langle E_F
\rangle-\frac{1}{60}\frac{\pi^2}{\langle E_F \rangle}T^2
\right)}_{E_{sym}^{kin}(n,T)}, \label{E-Kin-x}
\end{equation}
with the contribution of the symmetry energy  written explicitly.
It is obvious that in the low temperature limit
$E_{sym}^{kin}(n,T)$ behaves as $E_{sym}^{kin}(n,T)\propto C_1-C_2
T^2$.

\subsubsection*{High temperature limit}
The kinetic energy per nucleon $E_{kin}(n,T,\delta)$ of a
two-component Fermi gas at high temperature ($T\gg E_F$) is
replaced by a virial expansion in $n\lambda^3$ where
$\lambda=\sqrt{2\pi\hbar^2/mT}$ is the quantum wavelength. So,
$E_{kin}(n,T)$ is given by the relation \cite{Huang-87,Mekjian-05}
\begin{equation}
E_{kin}(n,\delta,T)= \frac{3}{2}T+\frac{3}{4}T\sum_\nu
C_{\nu}\left(\frac{\lambda^3n}{4}\right)^\nu\left((1-\delta)^{\nu+1}+(1+\delta)^{\nu+1}\right).
\label{Ek-3}
\end{equation}
Expanding expression (\ref{Ek-3}) around the symmetric point
$\delta=0$ or $x=1/2$ the kinetic energy takes the approximated
form
\begin{equation}
E_{kin}(n,T,\delta)=\frac{3}{2}T\left[1+ \sum_\nu
C_{\nu}\left(\frac{\lambda^3n}{4}\right)^\nu \right]+
(1-2x)^2\underbrace{\frac{3}{2}T \sum_\nu
C_{\nu}\left(\frac{\lambda^3n}{4}\right)^\nu\frac{\nu
(\nu+1)}{2}}_{E_{sym}^{kin}(n,T)}. \label{Ek-4}
\end{equation}
It is seen that in the high temperature limit $E_{sym}^{kin}(n,T)$
behaves as $E_{sym}^{kin}(n,T)\propto C_1T^{-1/2}+C_2
T^{-2}+\cdots$.

%\subsection{A1} The electron chemical potential $\mu_e$ could be
%calculated in the case of ideal Fermi gas at low temperature, in
%the beta stable matter, as follow. From beta equilibrium and
%neglecting muon we have
%\begin{equation}
%\mu_e=\mu_n-\mu_p \label{e-p-n-1}
%\end{equation}
%The chemical potential of neutron and proton when $T<<T_F$ takes
%the form
%\begin{equation}
%\mu_{\tau}=E_{F_{\tau}}\left[1-\frac{\pi^2}{12}\frac{k_BT^2}{E_{F_{\tau}}^2}\right],\qquad
%\tau=n,p \label{Chem-t-1}
%\end{equation}
%%
%Considering that
%\begin{equation}
%E_{F_{\tau}}=\frac{\hbar^2 k_{F_{\tau}}^2}{2m}=\frac{\hbar^2}{2m}
%(3\pi^2n_{\tau})^{2/3} \label{EF-1}
%\end{equation}
%and also that $n_n=(1-x)n$ and $n_p=xn$ after some algebra we get
%\begin{equation}
%\mu_{e}=\mu_0+\mu_T \label{Chem-t-2}
%\end{equation}
%where
%\begin{equation}
%\mu_0=\frac{20}{9}<E_F>(1-2x) \label{mu-o}
%\end{equation}
%and
%\begin{equation}
%\mu_T=\frac{\pi^2}{15}\frac{T^2 k_B^2}{<E_F>}(1-2x) \label{mu-T}
%\end{equation}
%%
%Considering in addition that the electrons are relativist that
%means
%\begin{equation}
%\mu_e=\hbar c (3\pi^2 n_e)^{1/3}=\hbar c (3\pi^2 n x)^{1/3}
\label{mu-e-1}
\section*{Acknowledgments}
%%%%%%%%%%%%%%%%%%%%%%%%%%
The author would like to thank Prof. S.E. Massen and Dr. C.P.
Panos for useful comments on the manuscript and also Prof. A.Z.
Mekjian for valuable comments and correspondence. The work was
supported by the Pythagoras II Research project (80861) of
E$\Pi$EAEK and the European Union.

%%%%%%%%BIBLIOGRAPHY%%%%%%%%%%%%%%%

\newpage

%%%FIGURES
%Fig-1
\begin{figure}
\centering
\includegraphics[height=6.0cm,width=5.5cm]{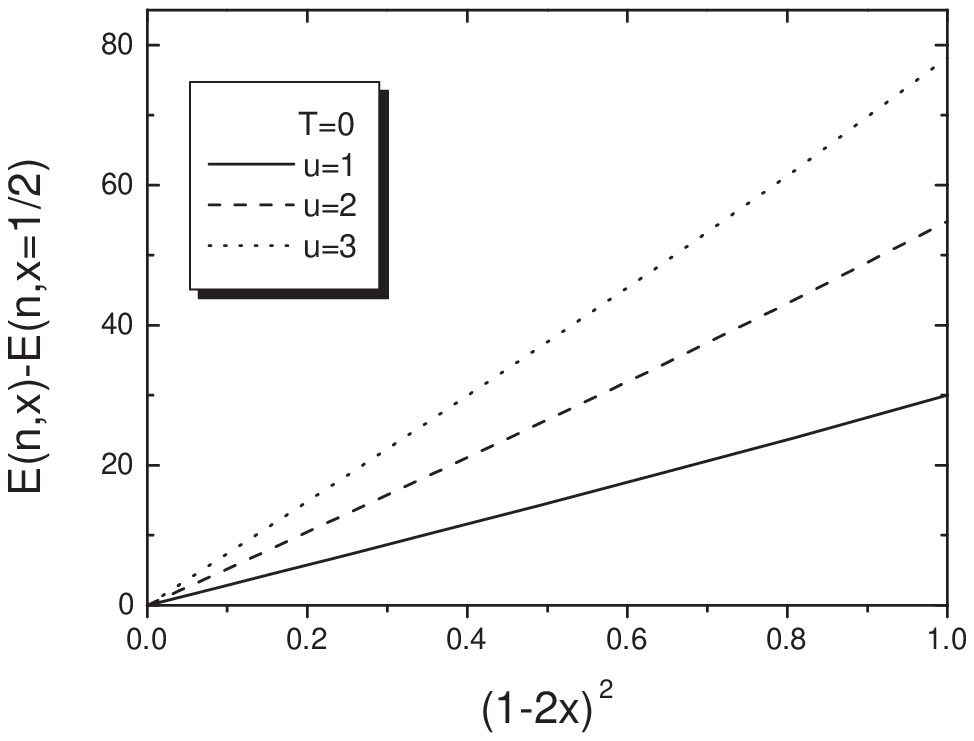}
\
 \includegraphics[height=6.0cm,width=5.5cm]{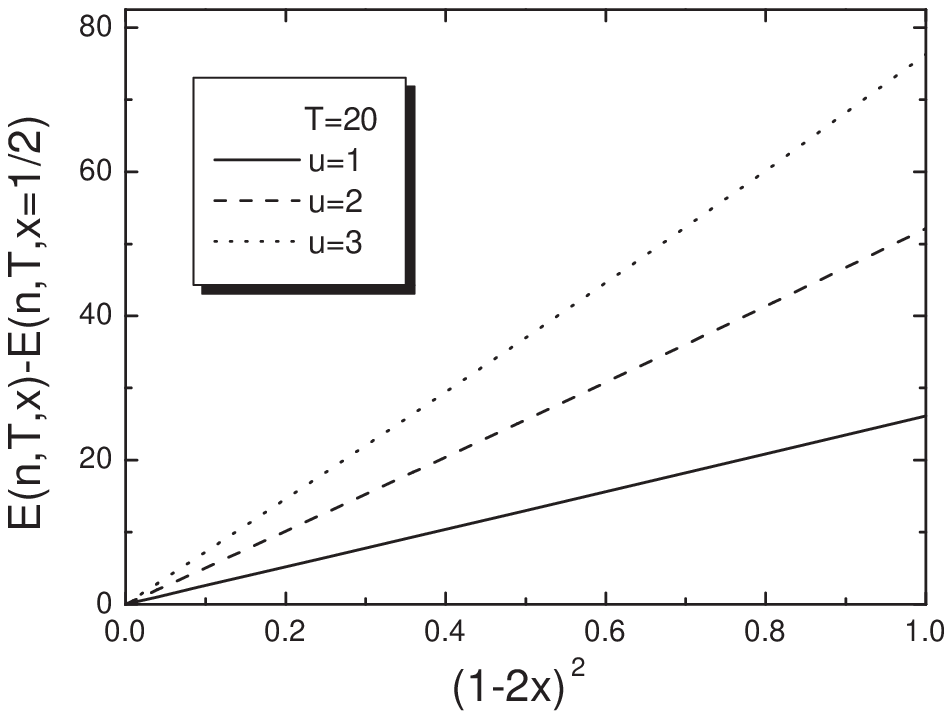}
\includegraphics[height=6.0cm,width=5.5cm]{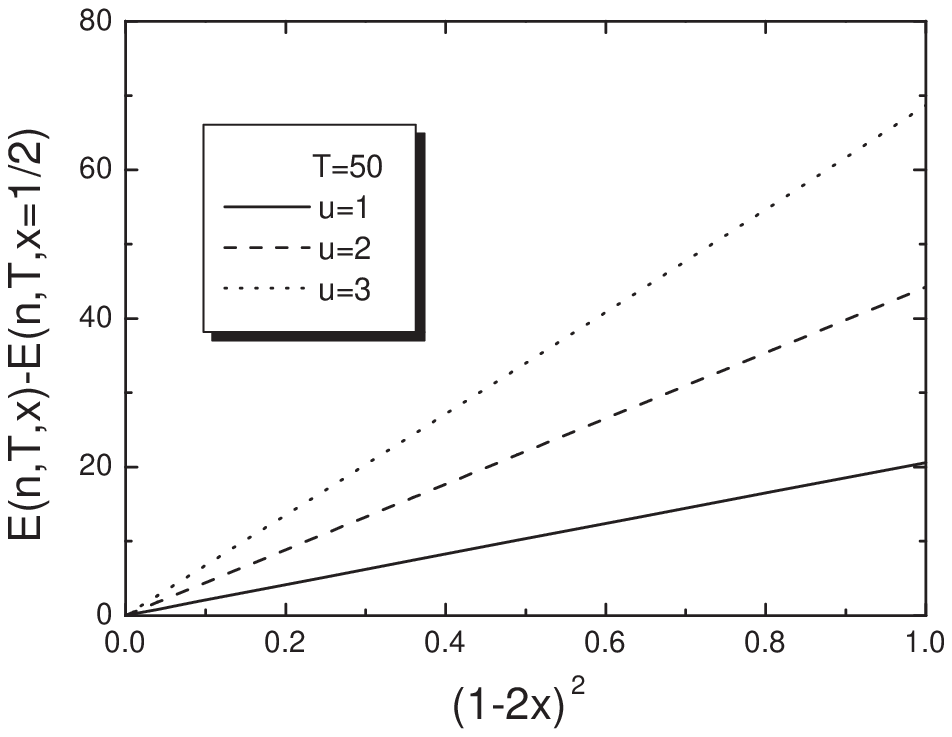}
\caption{The difference $E(n,T,x)-E(n,T,x=1/2)$ as a function of
$(1-2x)^2$ at temperature $T=0$, $T=20$ and $T=50$ MeV for three
baryon number fractions $u=1$, $u=2$ and $u=3$. } \label{}
\end{figure}
%%%%%%%%%%%%%%%%%%%%%%%%%%%%%%%%%%%%%%%%%%%%%%%%%%%%%%%%%%%%%%%%%%%%%%%%
%%%FIGURES
%Fig-2
\begin{figure}
\centering
\includegraphics[height=6.0cm,width=8.0cm]{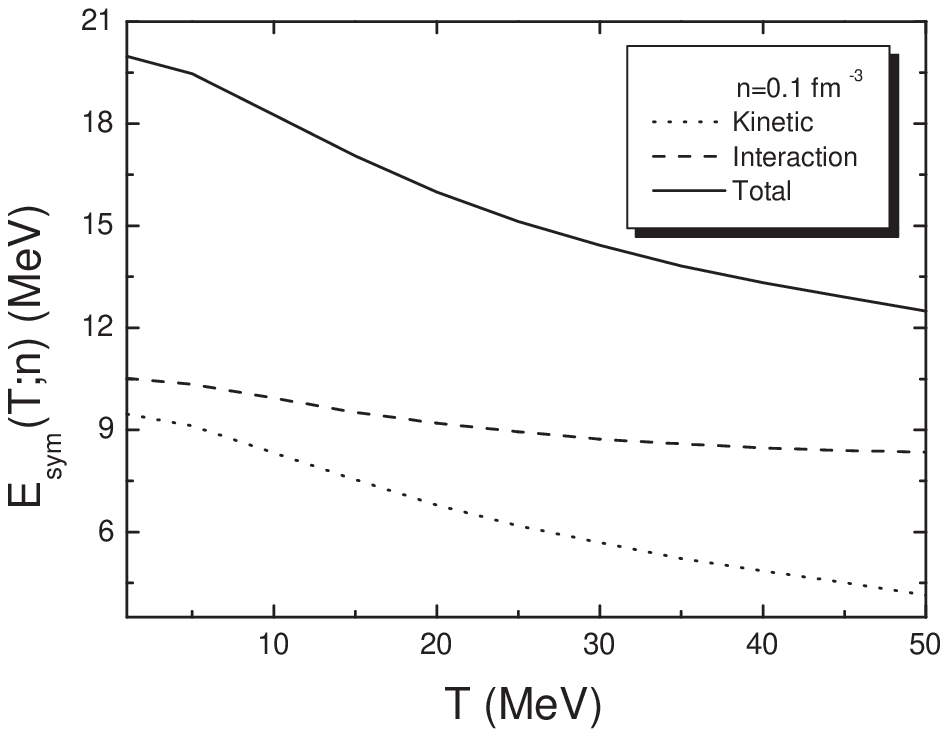}
\
 \includegraphics[height=6.0cm,width=8.0cm]{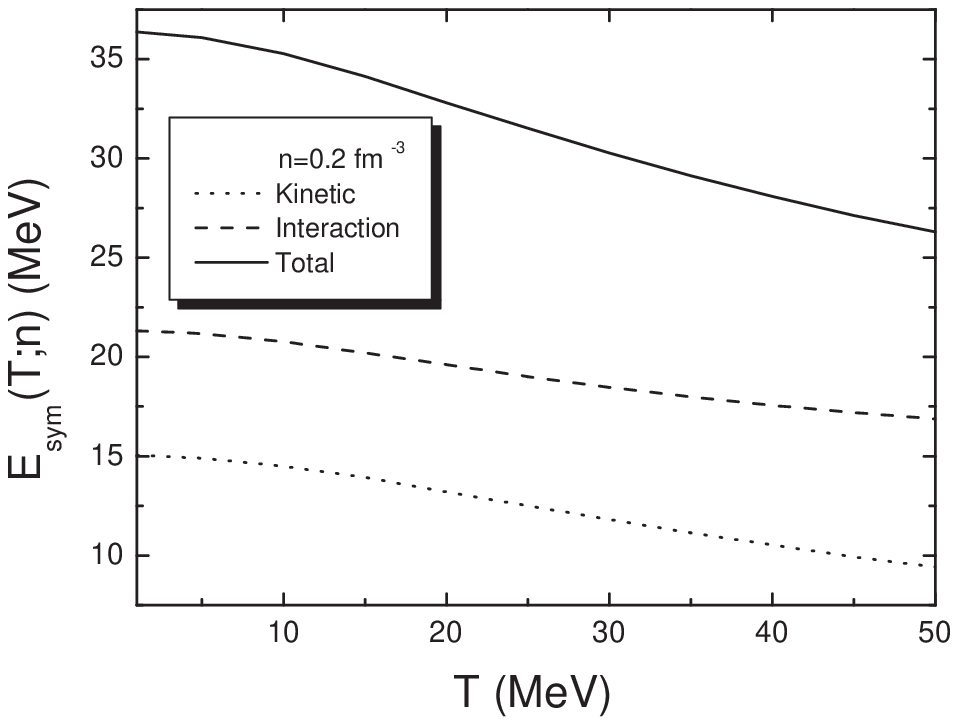}\\
\includegraphics[height=6.0cm,width=8.0cm]{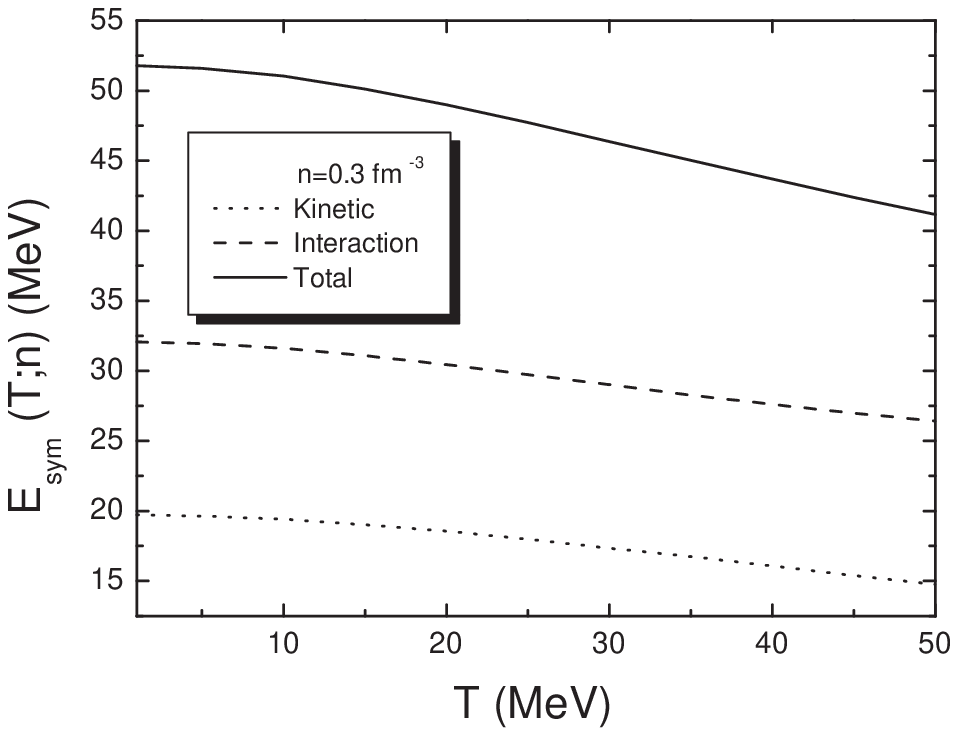}
\
 \includegraphics[height=6.0cm,width=8.0cm]{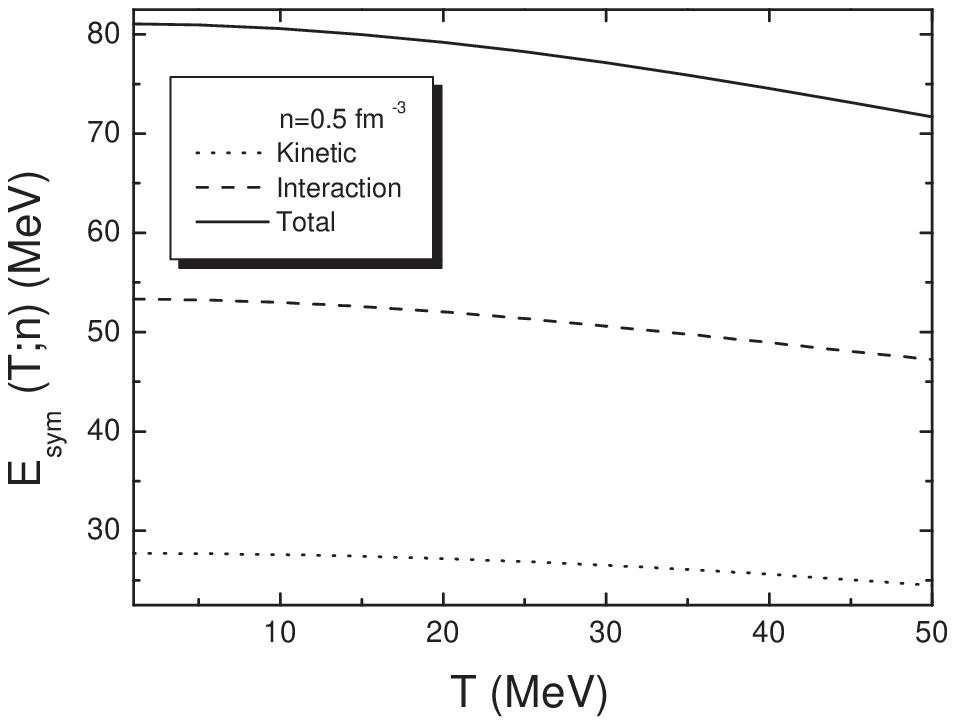}
\caption{Temperature dependence of the total nuclear symmetry
energy and its interaction and kinetic energy part for various
values of the baryon density $n$.} \label{}
\end{figure}
%%%%%%%%%%%%%%%%%%%%%%%%%%%%%%%%%%%%%%%%%%%%%%%%%%%%%%%%%%%%%%%%%%%%%%%%
\begin{figure}
\centering
\includegraphics[height=8.0cm,width=8.0cm]{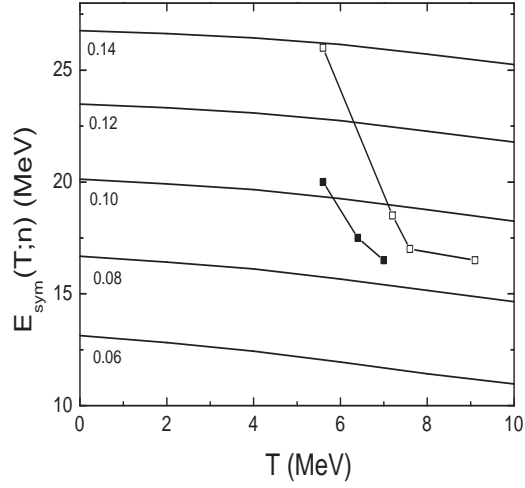}
\caption{Temperature dependence of the symmetry energy for low
values of the baryon density ($n=0.06, 0.08, 0.10, 0.12,0.14$
fm$^{-3}$). The experimental data are from Ref. \cite{Shetty-06}
(solid squares) and Ref. \cite{Fevre-05} (open squares) are
included for comparison. } \label{}
\end{figure}
%%%%%%%%%%%%%%%%%%%%%%%%%%%%%%%%%%%%%%%%%%%%%%%%%%%%%%%%%%%%%%%%%%%%%%%
%Fig-2
\begin{figure}
\centering
\includegraphics[height=8.0cm,width=5.0cm]{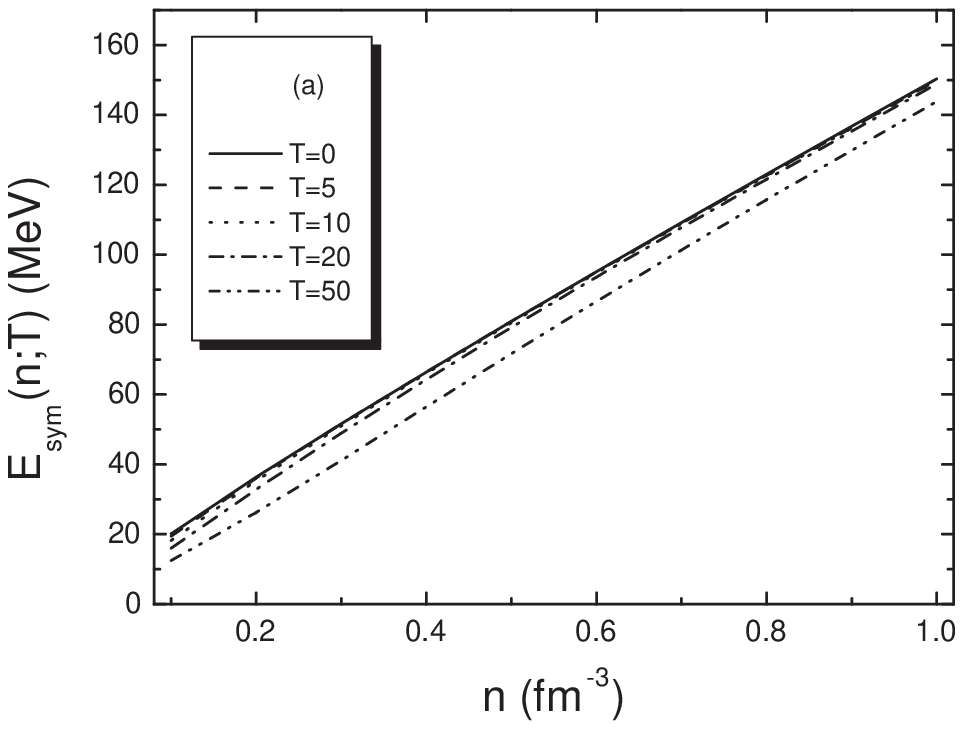}\
\includegraphics[height=8.0cm,width=5.0cm]{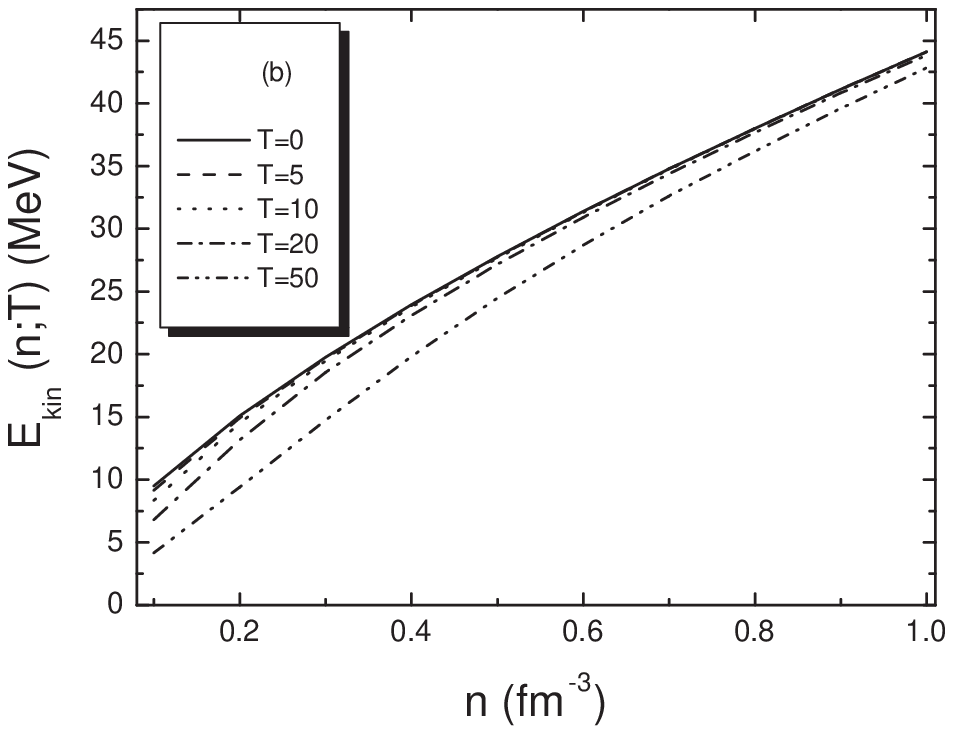}\
\includegraphics[height=8.0cm,width=5.0cm]{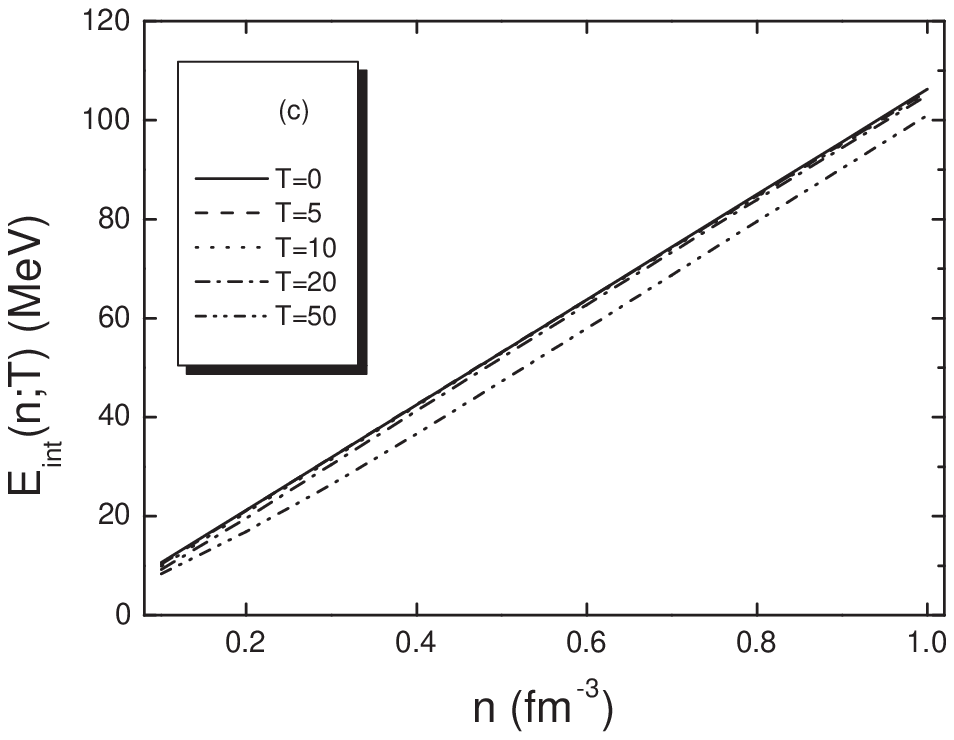}
\caption{Density dependence of the total nuclear symmetry energy
$E_{sym}^{tot}(n,T)$ as well its kinetic  $E_{sym}^{kin}(n,T)$ and
interaction $E_{sym}^{int}(n,T)$ part  for various values of the
temperature $T$.} \label{}
\end{figure}
%%%%%%%%%%%%%%%%%%%%%%%%%%%%%%%%%%%%%%%%%%%%%%%%%%%%%%%%%%%%%%%%%%%%%%%%%%%
%Fig-3
\begin{figure}
\centering
\includegraphics[height=9.0cm,width=7.0cm]{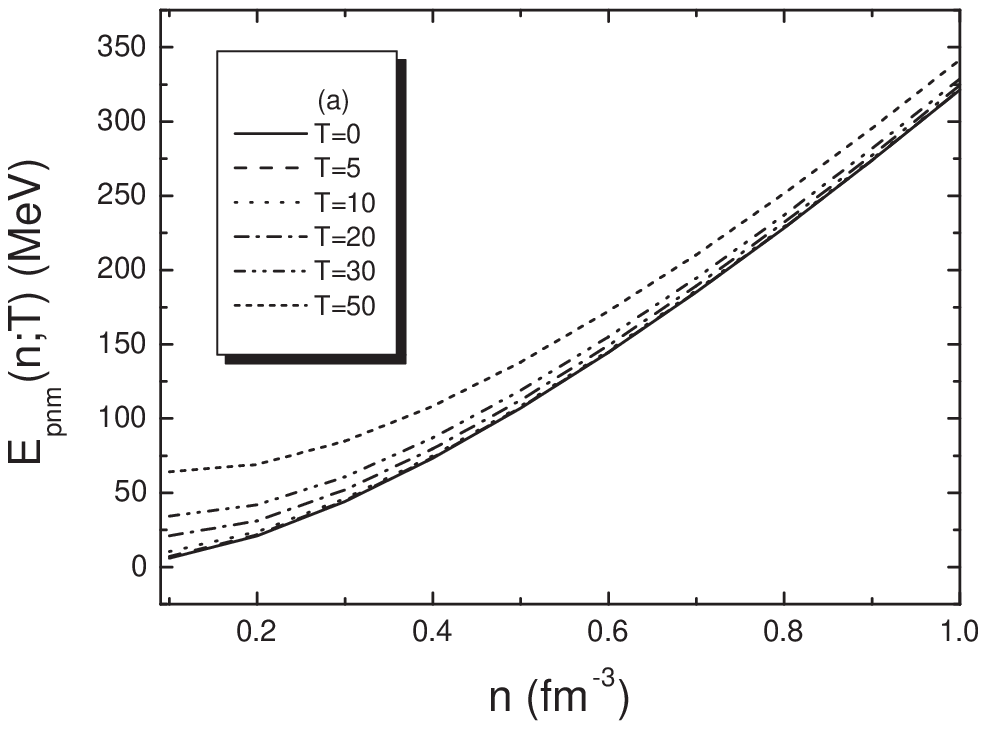}\
\includegraphics[height=9.0cm,width=7.0cm]{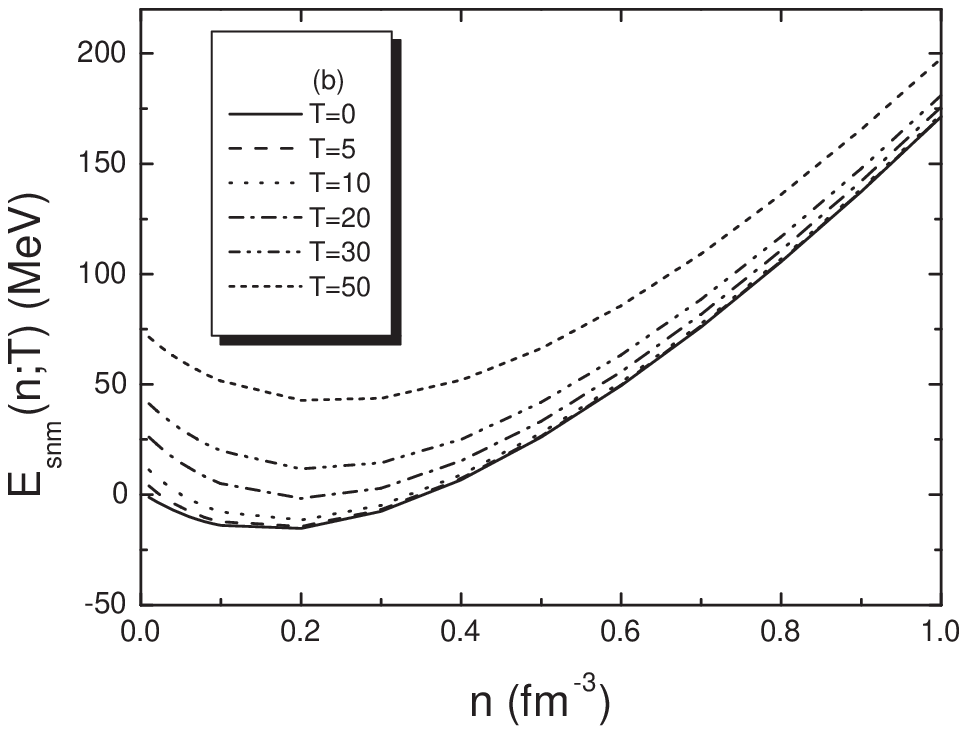}
\caption{(a) The energy per particle of pure neutron matter as a
function of the baryon density for various values of the
temperature $T$. (b) The energy per particle of symmetric nuclear
matter as a function of the baryon density for various values of
the temperature $T$. } \label{}
\end{figure}
%%%%%%%%%%%%%%%%%%%%%%%%%%%%%%%%%%%%%%%%%%%%%%%%%%%%%%%%%%%%%%%%%%%%%%%%
%Fig-4
\begin{figure}
\centering
\includegraphics[height=8.0cm,width=5.5cm]{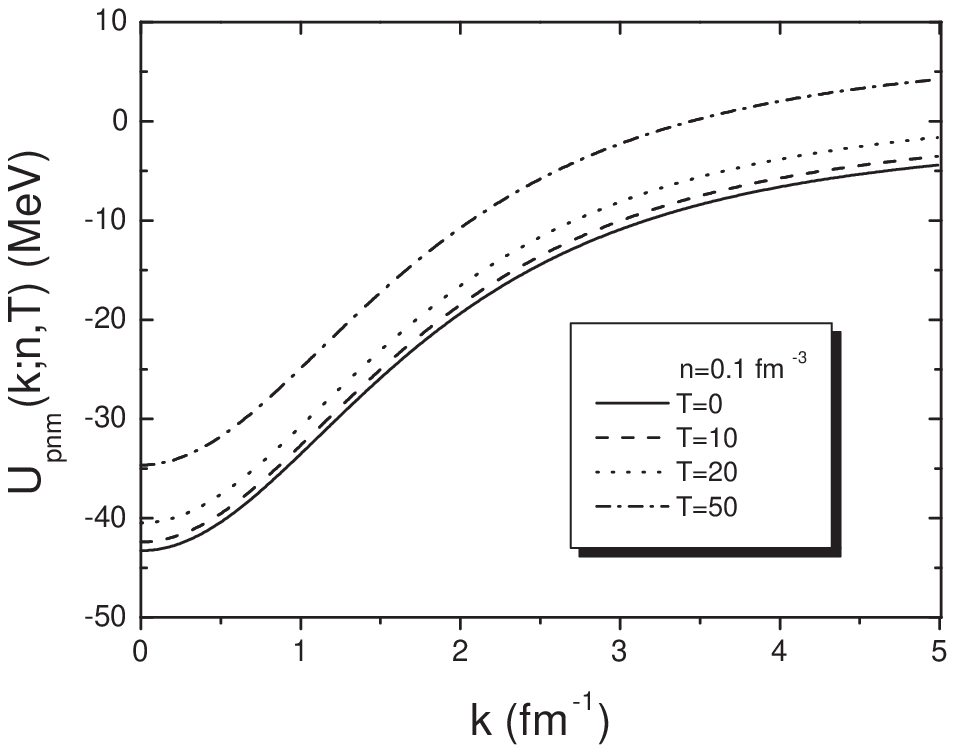}\
\includegraphics[height=8.0cm,width=5.5cm]{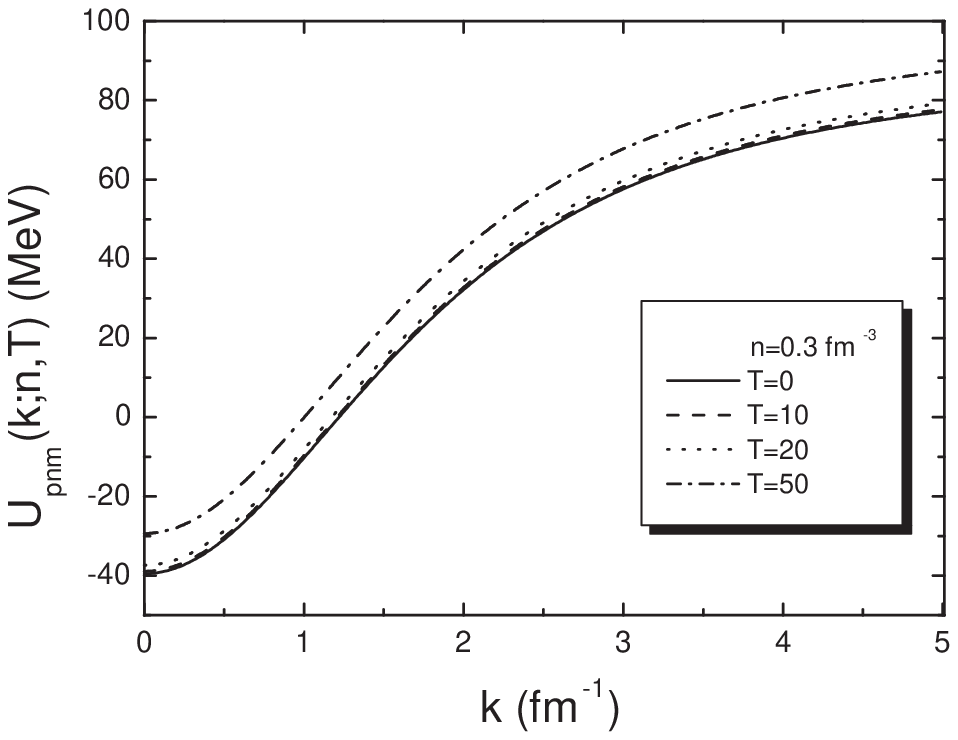}\
\includegraphics[height=8.0cm,width=5.5cm]{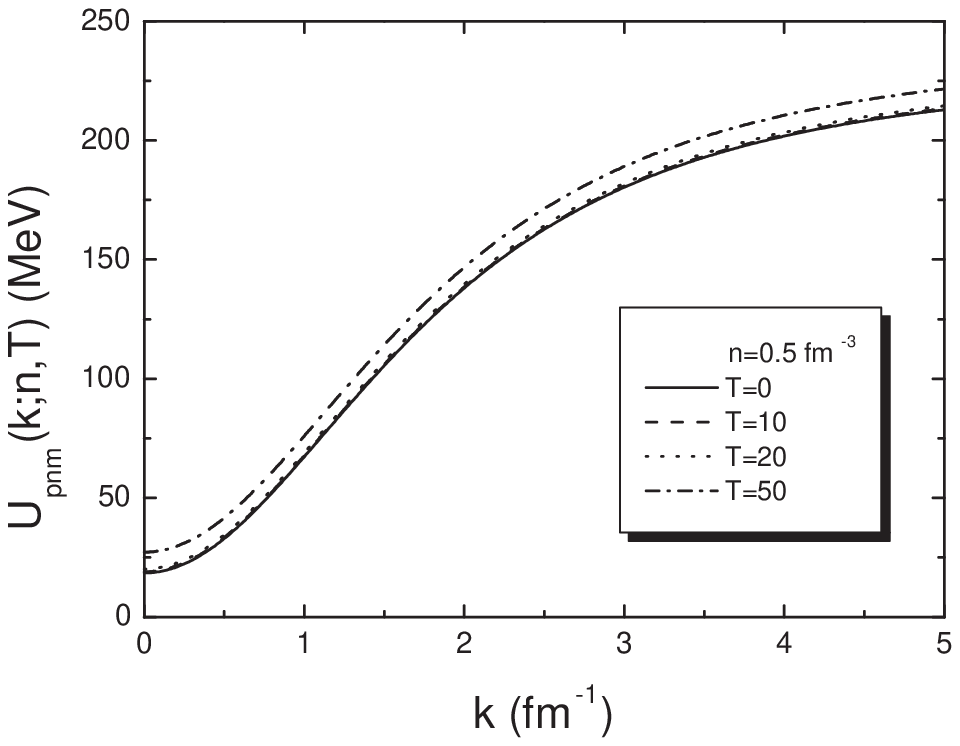}
\caption{The single particle potential of the pure neutron matter
as a function of the momentum $k$ for various values of the
temperature $T$ and for $n=0.1,0.3$ and  $0.5$ fm$^{-3}$
respectively.} \label{}
\end{figure}
%%%%%%%%%%%%%%%%%%%%%%%%%%%%%%%%%%%%%%%%%%%%%%%%%%%%%%%%%%%%%%%%%%%%%%%%
%Fig-5
\begin{figure}
\centering
\includegraphics[height=8.0cm,width=5.5cm]{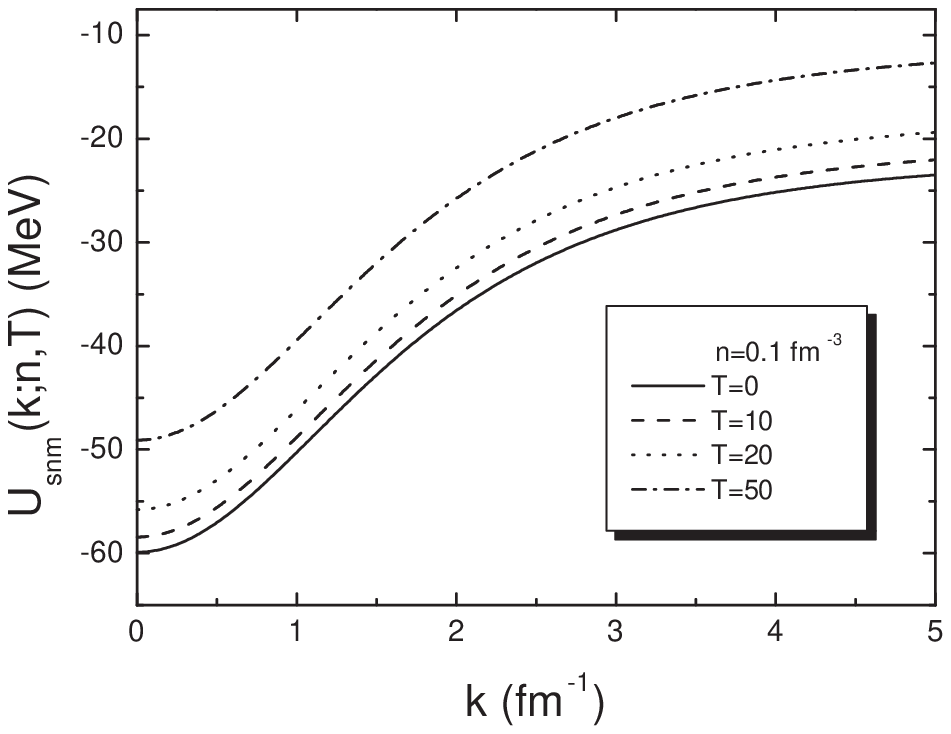}\
\includegraphics[height=8.0cm,width=5.5cm]{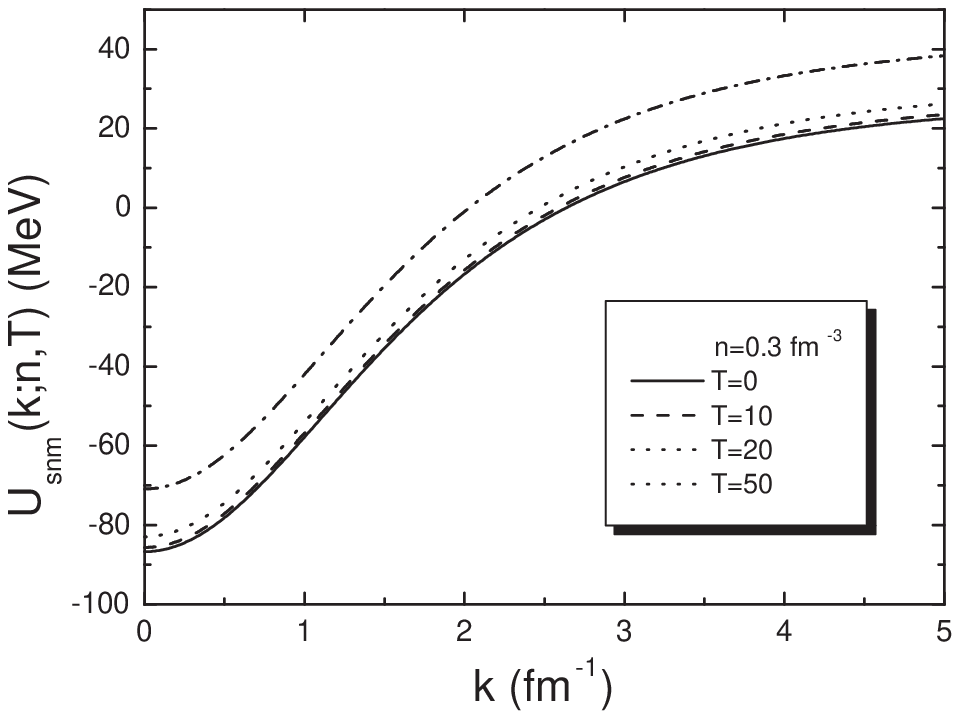}\
\includegraphics[height=8.0cm,width=5.5cm]{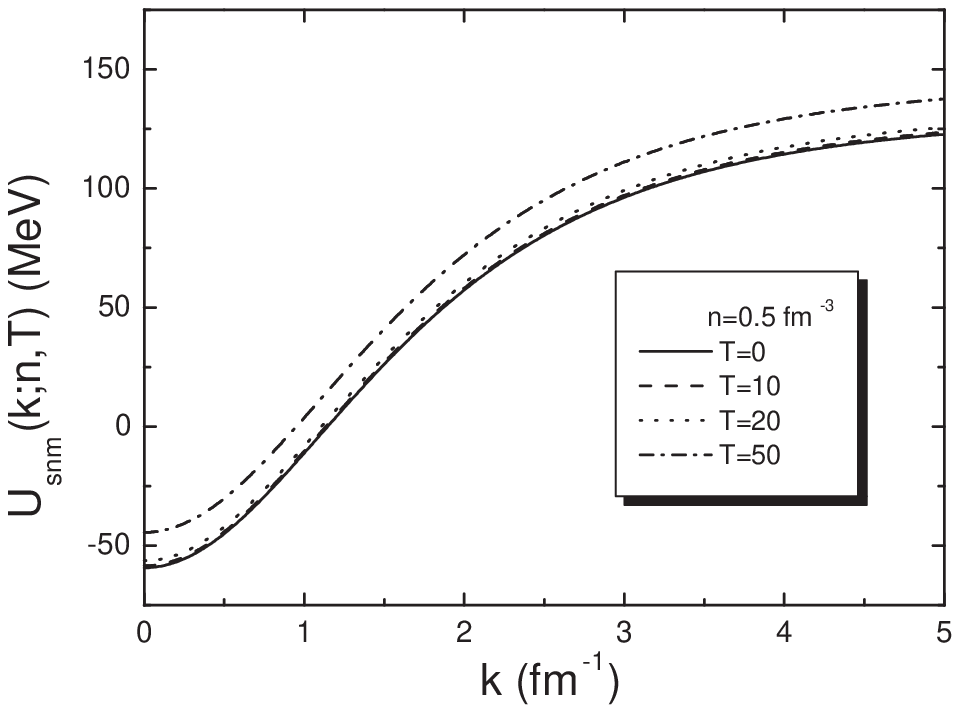}\
\caption{The single particle potential of the symmetric nuclear
matter as a function of the momentum $k$ for various values of the
temperature $T$ and for $n=0.1,0.3$ and  $0.5$ fm$^{-3}$
respectively.} \label{}
\end{figure}
%%%%%%%%%%%%%%%%%%%%%%%%%%%%%%%%%%%%%%%%%%%%%%%%%%%%%%%%%%%%%%%%%%%%%%%%%%%
%Fig-6
\begin{figure}
\centering
\includegraphics[height=8.0cm,width=5.5cm]{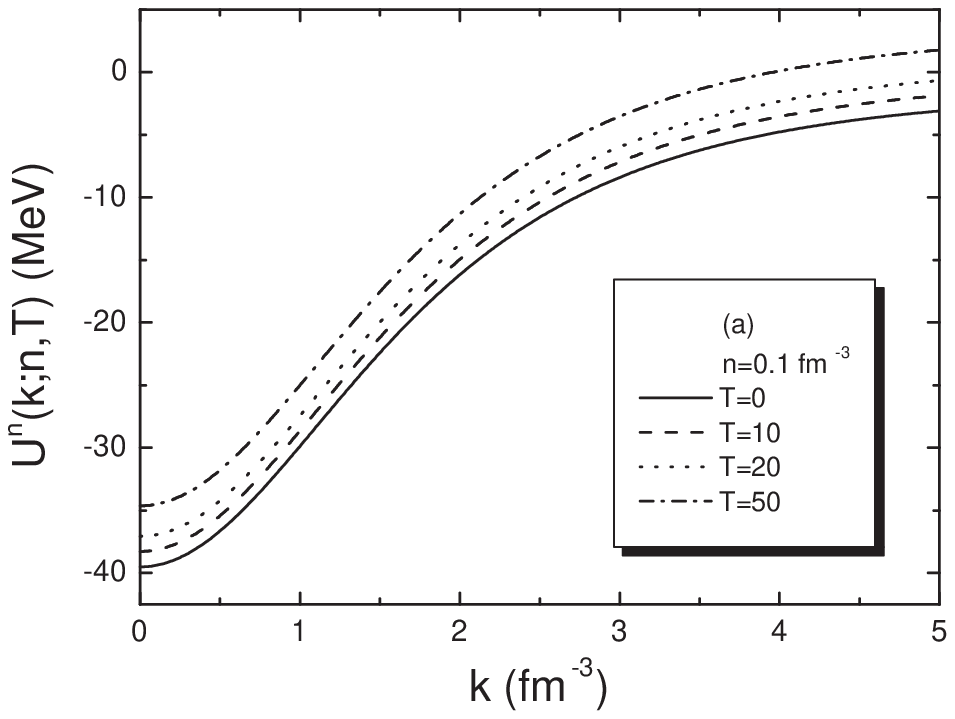}\
\includegraphics[height=8.0cm,width=5.5cm]{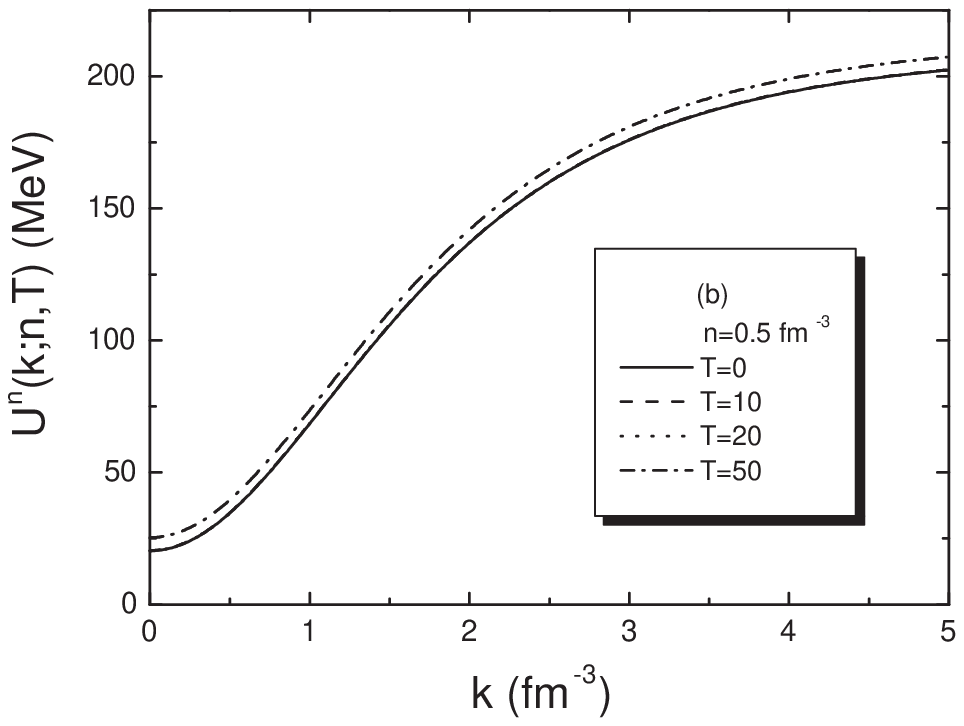}\\
\includegraphics[height=8.0cm,width=5.5cm]{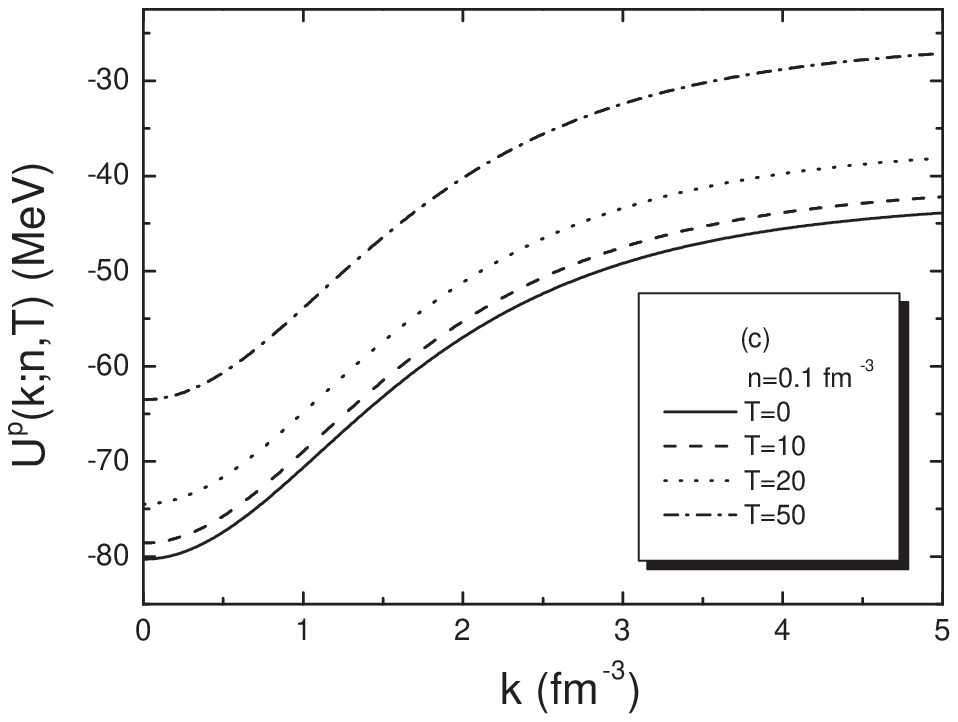}\
\includegraphics[height=8.0cm,width=5.5cm]{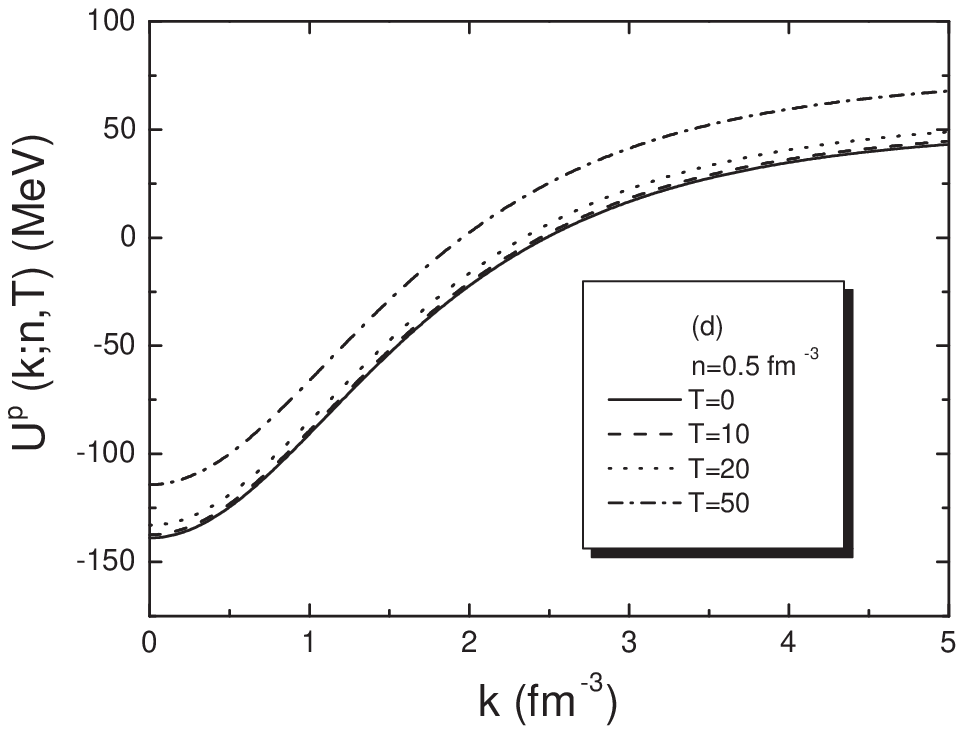}
\caption{The single particle potential of $\beta$-stable matter
for neutron ((a)and  (b)) and for proton ((c) and (d)) as a
function of the momentum $k$ for various values of the temperature
$T$  for $n=0.1$ and $0.5$ fm$^{-3}$ .} \label{}
\end{figure}

%%%%%%%%%%%%%%%%%%%%%%%%%%%%%%%%%%%%%%%%%%%%%%%%%%%%%%%%%%%%%%%%%%%%%%%%
%Fig-7
\begin{figure}
\centering
\includegraphics[height=8.0cm,width=8.0cm]{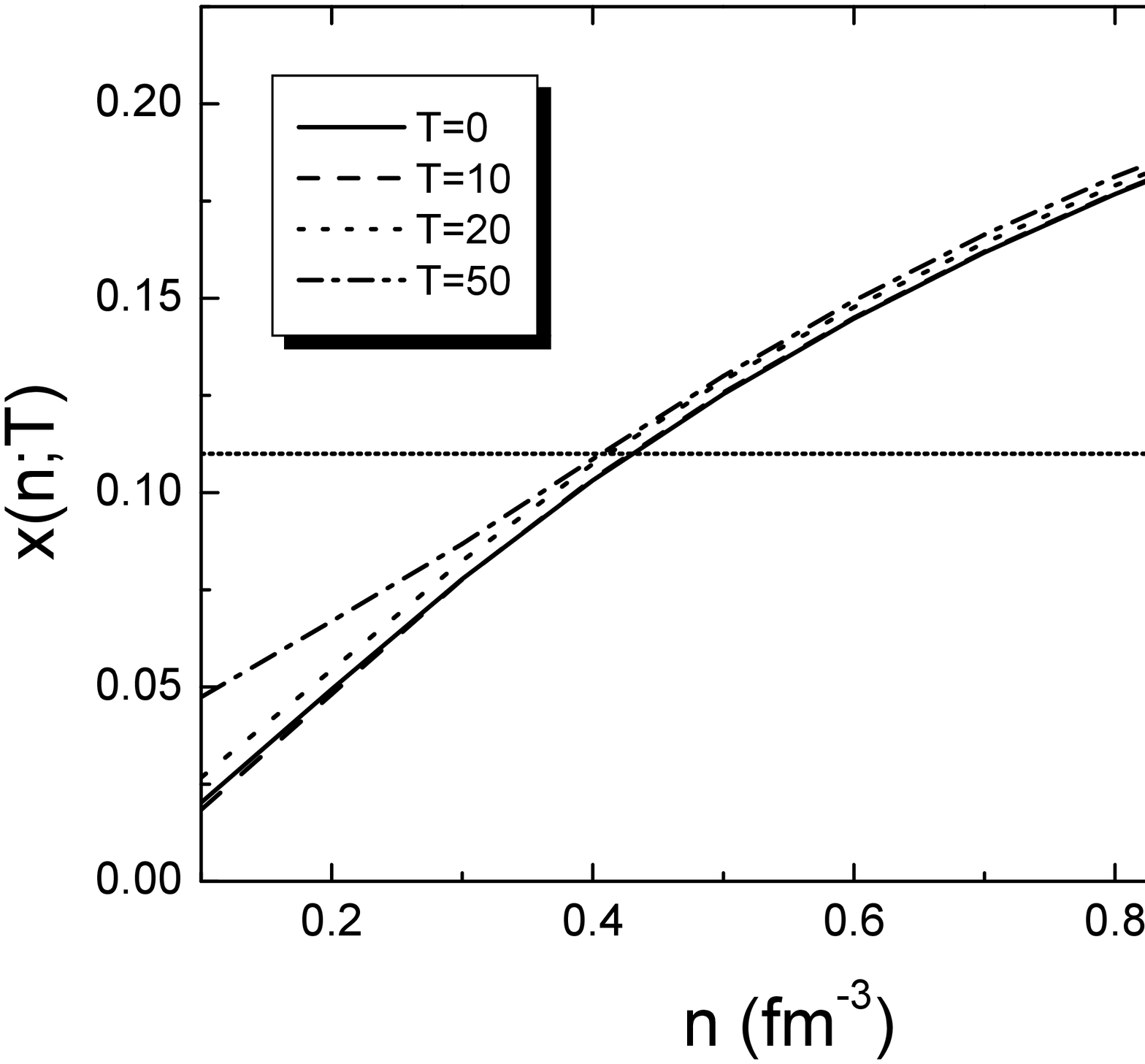}\
\includegraphics[height=8.0cm,width=8.0cm]{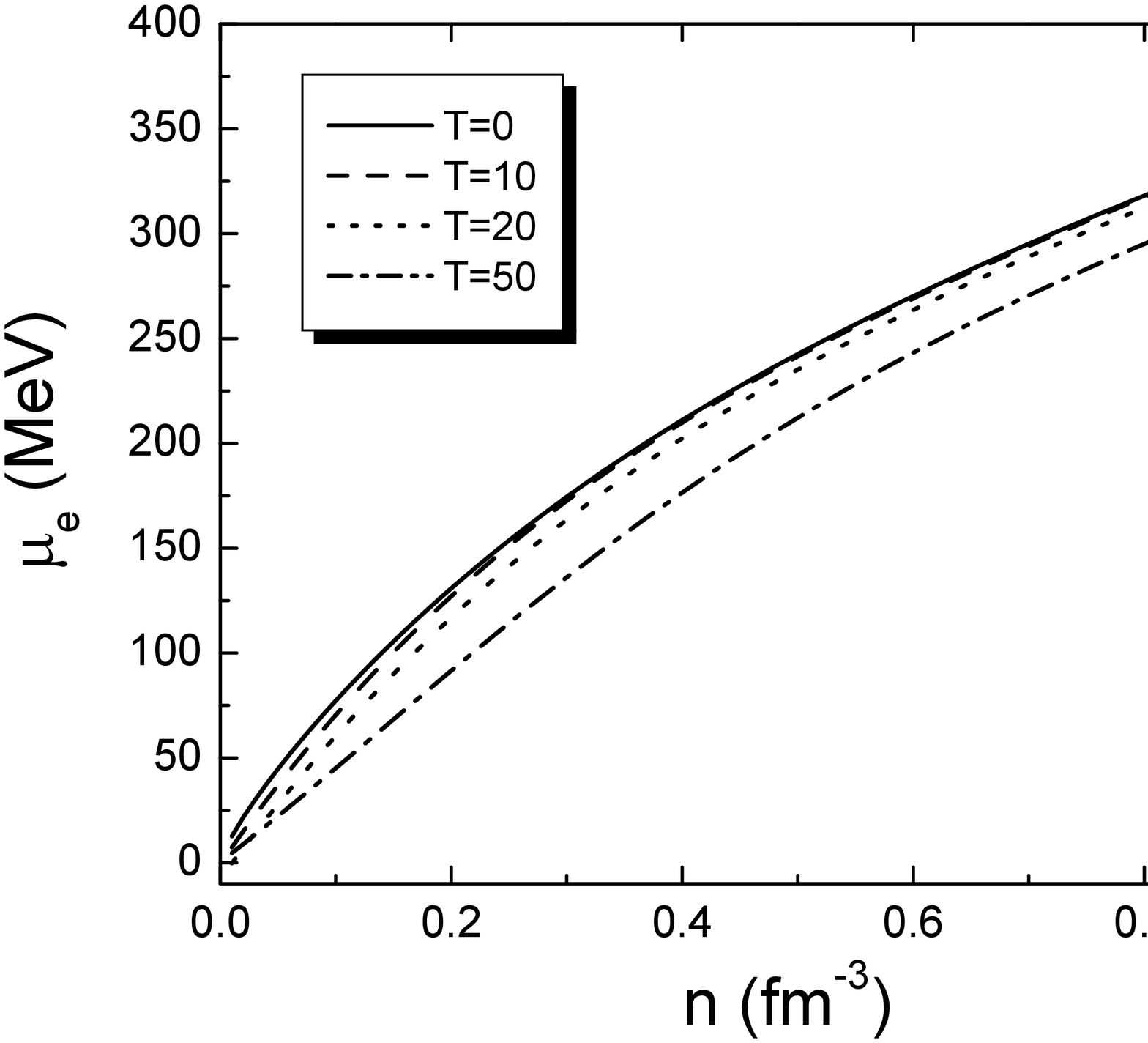}
\vspace{-3cm} \caption{(a) The proton fraction $x$ in
$\beta$-stable matter as a function of the density $n$ for various
values of the temperature $T$. The straight line corresponds to
the case $x= 11 \%$. (b) The electron chemical potential
$\mu_e=\hat{\mu}=\mu_n-\mu_p$ as a function of the density $n$ for
various values of the temperature $T$. } \label{}
\end{figure}
%%%%%%%%%%%%%%%%%%%%%%%%%%%%%%%%%%%%%%%%%%%%%%%%%%%%%%%%%%%%%%%%%%%%%%%%
%Fig-8
\begin{figure}
\centering
\includegraphics[height=8.0cm,width=8.0cm]{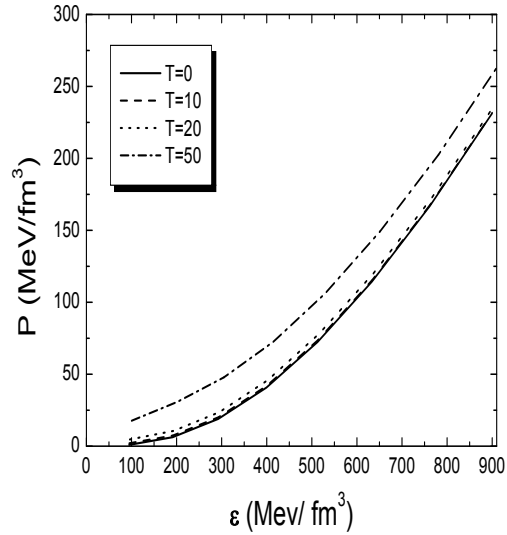}
\vspace{-3cm} \caption{The equation of state $P=P(\epsilon)$ of
$\beta$-stable matter corresponding to the present
momentum-dependent effective interaction model  for various values
of the temperature $T$.} \label{}
\end{figure}
%%%%%%%%%%%%%%%%%%%%%%%%%%%%%%%%%%%%%%%%%%%%%%%%%%%%%%%%%%%%%%%%%%%%%%%%
\newpage
%%%%%
 \begin{table}[h]
\begin{center}
\caption{The values of the density dependent parameters $A$, $B$,
$T_0$ and $c$, for $E_{sym}^{tot}(u;T)$, $E_{sym}^{kin}(u;T)$ and
$E_{sym}^{int}(u;T)$ for $n=0.1, 0.3, 0.5$ fm$^{-3}$. For more
details see text. }
 \label{t:1}
\vspace{0.5cm}
\begin{tabular}{|ccccccccccc|}
\hline
 &    & n=0.1 fm$^{-3}$  &    &  & n=0.3 fm$^{-3}$  &    &  & n=0.5 fm$^{-3}$  & &      \\
%\hline
 ${\rm Parameters}$   & $E_{sym}^{tot}$ &
$E_{sym}^{kin} $ & $E_{sym}^{int}$ & $E_{sym}^{tot} $ &
$E_{sym}^{kin}$ & $E_{sym}^{int}$ & $E_{sym}^{tot}$ & $E_{sym}^{kin}$ &  $E_{sym}^{int}$&   \\
\hline
 $A$  & 10.105  & 7.864  & 2.559 & 17.079 & 15.230  & 10.832  & 44.164 & 20.548 & 19.442& \\
 \hline
 $B$  & 9.969     & 1.679  & 7.969  & 19.328 & 4.504    & 21.240  & 36.895 & 7.162  &33.887& \\
 \hline
 $T_0$  & 25.692   & 30.549  & 19.027 & 41.004 & 73.143  &  47.772 & 57.011 &109.193  & 73.551& \\
 \hline
 $c$  & 1.610   & 1.518  & 1.866 & 1.856 & 1.904  &  1.992 & 1.982 & 2.156   &  2.026& \\
 \hline
 %\hline
\end{tabular}
\end{center}
\end{table}

\end{document}